\documentclass[comunication,amsfonts,amssymb,twocolumn]{revtex4}
\usepackage{amsfonts}
\usepackage{graphicx,hyperref}
\usepackage{amsmath,graphicx}
 \allowdisplaybreaks

\usepackage{bm}
\usepackage{CJK}
\begin{document}

\title{Dynamics of spin-orbit-coupled cold atomic gases in a Floquet lattice with an impurity}
\author{Xiaobing Luo$^{1,2,3}$}
\altaffiliation{Author to whom any correspondence should be addressed: xiaobingluo2013@aliyun.com}
\author{Baiyuan Yang$^{1,2}$}
\author{Jin Cui$^{3}$}
\author{Yu Guo$^{4}$}
\author{Lei Li$^{1,2}$}
\author{Qianglin Hu$^{1,2}$}
\affiliation{$^{1}$ Department of Physics, Jinggangshan University,
Ji'an 343009, China}
\affiliation{$^{2}$ Institute of Atomic, Molecular Physics $\&$ Functional Materials, Jinggangshan University, Ji'an 343009, China}
\affiliation{$^{3}$Department of Physics and Key Laboratory of
Low-dimensional Quantum Structures and Quantum Control of Ministry
of Education, \\
and Synergetic Innovation Center for Quantum Effects and Applications, Hunan Normal University, Changsha 410081, China}
\affiliation{$^{4}$ School of Physics and Electronic Science, Changsha University of Science and Technology, Changsha  410114, China}
\date{\today}
\begin{abstract}
In this study, we have studied the quantum tunneling of a single spin-orbit-coupled atom held in a periodically modulated optical lattice with an impurity.
At the pseudocollapse points of quasienergy bands, where the dynamical localization takes place globally, two types of local second-order tunneling processes appear beyond expectation between the two
nearest-neighbor sites of the impurity with the spin unchanged and with impurity site population negligible all the time, when the impurity potential is far off-resonant with the driving field. Though  tunneling behaviors of the two types seem to be the same, they are believed to involve two distinct mechanisms: one is related to spin-independent process, while the other is to spin-dependent tunneling process. The two types of second-order processes can be identified by means of resonant tunneling with or without spin-flipping by tuning
the impurity potential to be in resonance with the driving field. In the Floquet picture, the second-order processes are manifested as subtle and fine avoided crossings of quasienergy spectrums near the pseudocollapse region. These results are confirmed analytically on the basis of effective
three-site model and multiple-time-scale asymptotic perturbative method, and may be exploited for  engineering the spin-dependent quantum transport in realistic experiments.
\end{abstract}

\maketitle
\section{Introduction}
Spin-orbit coupling (SOC) effect, ubiquitous in condensed matter physics, lies at the heart of fundamental
phenomena such as spin Hall effect\cite{Sinova}, topological
insulators\cite{Qi}, Majorana
fermions\cite{Sau}, as well
as practical applications such as spintronic devices\cite{Koralek}.
Recently, artificial spin-orbit coupling (SOC) has been successfully engineered in the laboratory with both neutral bosonic and fermionic
ultracold atoms\cite{Lin,Pan1,Wang,Cheuk,Huang,Pan}, which not only exhibits many exotic phases\cite{Zhai,Wu,Stanescu,Sinha,Zhai2,Ho,Li,Hu} (some of which
have no direct analog in conventional condensed
matter systems), but also provides an ideal platform to recreate and explore novel
SOC physics with an unprecedented level of tunability of experimental parameters.
Motivated by the ongoing
experimental achievements, scholars have put extensive
efforts toward understanding the
static properties of the SO-coupled
atomic gases\cite{Li,Hu,Wen,Zhang,Cole,Radic,Gong,Cai,Zhu1,Xu}, as well as gaining insight into their intriguing dynamics, including unconventional
collective dipole oscillations\cite{Pan3,Zhai3}, relativistic dynamics with analogs
of Zitterbewegung\cite{Qu} and Klein tunneling\cite{zhu2}, spin Josephson effects in a double-well potential\cite{zhu3,Garcia-March,Citro}, spin-dependent
tunneling and spin transportation in Floquet systems of SO-coupled ultracold atoms\cite{Xue,Hai1}, nonequilibrium dynamics of
SO-coupled lattice bosons\cite{Ng}, tunable Landau-Zener transitions\cite{Olson} and chaos-driven dynamics\cite{Larson} in SO-coupled atomic gases, localization of a SO-coupled  particle or Bose-Einstein condensate moving
in a 1D quasiperiodic potential\cite{LZhou,Adhikari} and random potential\cite{Edmonds,Mardonov}, entanglement\cite{Hai2,Sun} and dynamical phase transition\cite{Sun} in a SO-coupled Bose-Einstein condensate, and so on.

When SOC is combined with optical lattices, a
spin-flip term is produced in Bose-Hubbard Models, in addition to the standard spin-conserving
tunneling between nearest-neighbor lattice sites.
In recent experiments, SO-coupled bosonic gases in an optical lattice have been successfully realized\cite{Hamner}, which promises a testbed for theoretical
studies of diverse new phases and SOC-related spin-dependent transportations.
Typically, defects play a prominent
role in the dynamical properties of the realistic lattice systems.
A large number of papers have been devoted to theoretical investigation of the behaviors of quantum particles without SOC hopping on a defective lattice,
such as the existences of bound state in
the continuum (BIC)\cite{JMZhang} and Floquet bound states\cite{Zhong} in the one-dimensional two-particle Hubbard model with an impurity,
the resonant tunneling of a single particle\cite{Liang} and correlated pair tunneling\cite{Longhi,Zhou1} in
driven one-dimensional lattices with an impurity, to name only a few.
By contrast, the physics of
a single SO-coupled ultracold atom in a driven one-dimensional optical lattice with an impurity
remains an unexplored frontier. On the other hand, for controlling more favorably
 spin-dependent
tunneling and spin transportation in lattices,
 disorder and external driving field may be intentionally introduced. From this point of view,
it would be
worthwhile to study the dynamics of a SO-coupled particle moving in a periodically driven lattice with an impurity.

This paper is principally aimed  to understand how the interplay between impurity, SOC and periodic driving determines the novel tunneling dynamics of a single SO-coupled atom moving in an optical lattice. As we find it, the dynamical localization (DL) is not destroyed when considering tunneling dynamics of a  single SO-coupled atom with only spin-conserving
coupling or with only spin-flipping
coupling in the driven optical lattice with an impurity, but the local dynamics of the system is remarkably affected when the competition among impurity, SOC and periodic driving is introduced.
Specifically, there are two scenarios.
In the resonant regime, where the impurity potential is equal to a multiple of the driving frequency, the dynamics of the system is dominated by resonant tunneling between the impurity and its two nearest-neighbor sites. Such a system with only coupling between
states with the same spin will exhibit resonant tunneling without
spin flipping, while the one with only coupling between states with different
spins demonstrates resonant tunneling with spin flipping.
In the non-resonant regime, where the impurity potential is far off-resonant with
the driving field,
 the second-order tunneling process emerges surprisingly even under the DL conditions when either only the spin-conserving tunneling term or only the spin-flipping tunneling term is presented; in either case, a single spin-up (down) atom
tunnels back and forth between the two nearest-neighbor
sites of the impurity in such a way that the spin remains unchanged and the impurity site population is
negligible over all the evolution time. For the single SO-coupled atom with either only spin-conserving
coupling or with only spin-flipping
coupling term, the second-order tunneling processes
exhibit remarkable resemblance, though, they are generated via different virtual intermediate states and therefore should be interpreted in the context of two distinct mechanisms.
The first mechanism involves a spin-independent tunneling process in which the virtual intermediate state without spin-flipping is eliminated, thereby yielding a spin-independent effective tunneling rate. The second one stems from a second-order transition via elimination of the virtual intermediate state with spin-flipping and actually involves a spin-dependent tunneling process, thus second-order transition of this type has a spin-dependent effective tunneling rate.

\section{Model system and Floquet quasienergy spectrum}
We consider a single ultracold atom with internal up and down pseudospin
states $|\uparrow\rangle$ and $|\downarrow\rangle$  trapped in
 a spin-independent one-dimensional periodically driven optical lattice
with an impurity at site $n=0$. The effective spin-orbit coupling can be implemented
simultaneously experimentally using two counter-propagating Raman lasers which generate a momentum-sensitive
coupling between the two internal hyperfine states of the same
atom. In the most general tight-binding description, the single-particle Hamiltonian governing this system is of the following
form\cite{Xue,Hai1,LZhou}
\begin{eqnarray} \label{Hamiltonian}
 H&=&\sum\limits_{n}\large [-(\hat{c}_{n}^{\dagger}\hat{T}\hat{c}_{n+1}+h.c.)+\varepsilon'(t) n\hat{c}_{n}^{\dagger}\hat{c}_{n}
+\frac{\Omega}{2}\hat{c}_{n}^{\dagger}\hat{\sigma}_{z}\hat{c}_{n}\large ]\nonumber\\
&&+\varepsilon_{0}\hat{c}_{0}^{\dagger}\hat{c}_{0},
\end{eqnarray}
where $\hat{c}^{\dagger}_{n}=(\hat{c}^{\dagger}_{n\uparrow},\hat{c}^{\dagger}_{n\downarrow}), \hat{c}_{n}=(\hat{c}_{n\uparrow},\hat{c}_{n\downarrow})^{T}$ (hereafter superscript $T$ stands for the transpose) and
$\hat{c}^{\dagger}_{n\sigma} (\hat{c}_{n\sigma} )$ creates (annihilates) a boson with pseudo-spin $\sigma=\uparrow,\downarrow$
at site $n$ ($n=0,\pm 1,\pm 2,...$),  $\hat{\sigma}_{x,y,z}$ are the usual $2\times2$ Pauli matrices, $\Omega$ is the effective
Zeeman field intensity, and $\varepsilon_{0}$ the impurity potential strength. $\hat{T}\equiv v\exp(-i\alpha\hat{\sigma}_{y})=v(\cos\alpha-i\hat{\sigma}_{y}\sin\alpha)$ is the hopping matrix obtained by the Peierls substitution, where $v$ represents the hopping amplitude in the absence of synthetic SO coupling, and the dimensionless parameter $\alpha=\pi k_{r}/k_{\rm{lat}}$ characterizes the
strength of SO coupling with $k_{r}$ and $k_{\rm{lat}}$
being the wave vector of the Raman laser and the lattice respectively. For convenience, we will refer to dressed pseudo-spin as spin hereafter.
The diagonal
terms of $\hat{T}$ describe spin-conserving hopping while off-diagonal spin-flip
terms describe the SO coupling arising from a two-photon
Raman process. In our model, we assume that the driving
field take the form $\varepsilon'(t)=F\cos(\omega t)$  with amplitude $F$ and
frequency $\omega$.
In the numerical simulations
below, we take the size of lattice to be $N=21$, i.e., $n=0,\pm 1,\pm 2,...,\pm 10$, unless explicitly stated otherwise.

For simplicity, we have set $\hbar=1$ and adopted the parameter $v$ as
a unit to scale the other parameters $\Omega, F, \omega, \varepsilon_0$ so that they become dimensionless.
In realistic experiment, the Zeeman field $\Omega$ is on order of recoil frequency $E_r/\hbar=\hbar k_r^2/(2m)=22.5\rm{kHz}$\cite{Lin},
$v\sim 0.1E_r/\hbar$, and the driving frequency $\omega$ can be adjusted between $0\sim 30 \rm{kHz}$\cite{Bloch}.
Thus, the system parameters can be tuned experimentally
in a wide range as follows $\Omega\sim 10v, F\sim\varepsilon_0\sim\omega\in[0,10]v$. In our discussion, what is essential
is the ratios between these parameters $v,\Omega, F, \omega, \varepsilon_0$.

In Hilbert space with a complete set of Fock basis $\{|n ,\sigma\rangle\}$, where $|n,\sigma\rangle$ represents the state of a spin-$\sigma$ ($\sigma=\uparrow,\downarrow$) particle occupying a lattice site $n$, the state vector of the SO-coupled system at any time $t$ can be expanded as
\begin{eqnarray}\label{Statevector}
    |\psi(t)\rangle=\sum\limits_{n,\sigma}a_{n,\sigma}(t){|n ,\sigma\rangle},
\end{eqnarray}
where $a_{n,\sigma}(t)$ indicates the time-dependent probability amplitude of the atom being in state $|n,\sigma\rangle$, and the corresponding
probabilities read $P_{n ,\sigma}=|a_{n,\sigma}(t)|^2$, conserving the normalization condition $\sum_{n,\sigma}P_{n ,\sigma}=1$.
Substituting Eqs.~(\ref{Hamiltonian}) and (\ref{Statevector}) into Schr$\ddot{\textrm{o}}$dinger equation $i\partial_{t}|\psi(t)\rangle=H|\psi(t)\rangle$, one obtains the following coupled equations for the probability amplitude $a_{n,\sigma}$
\begin{eqnarray}\label{AmPro}
   i\dot{a}_{n,\uparrow}&=&-\upsilon[\cos\alpha({a}_{n+1,\uparrow}+{a}_{n-1,\uparrow})
   \nonumber\\&&+\sin\alpha(-{a}_{n+1,\downarrow}+{a}_{n-1,\downarrow})]\nonumber\\
   &&+\varepsilon'(t)n{a}_{n,\uparrow}+\varepsilon_{0}\delta_{n,0}{a}_{n,\uparrow}+\frac{\Omega}{2}{a}_{n,\uparrow}, \\   i\dot{a}_{n,\downarrow}&=&-\upsilon[\cos\alpha({a}_{n+1,\downarrow}+{a}_{n-1,\downarrow})\nonumber\\&&+\sin\alpha({a}_{n+1,\uparrow}-{a}_{n-1,\uparrow})]
   \nonumber\\
   &&+\varepsilon'(t) n{a}_{n,\downarrow}+\varepsilon_{0}\delta_{n,0}{a}_{n,\downarrow}-\frac{\Omega}{2}{a}_{n,\downarrow}.\label{AmPro2}
\end{eqnarray}
Throughout our paper, the overdot represents a derivative with respect to the time variable $t$, unless stated otherwise. From Eqs.~(\ref{AmPro}) and (\ref{AmPro2}), it is easy to observe that the couplings between states with the same (different) spin are
proportional to cosine (sine) functions of the SOC strength.
The terms proportional to $\cos\alpha$
are the usual spin-conserving tunneling, while those
proportional to $\sin\alpha$  are the
spin-flipping tunnelling. When
$\alpha=l\pi,l=0,1,2,....$, only the usual tunneling is presented;
When $\alpha=(2l+1)\pi/2,l=0,1,2,....$, the usual tunneling vanishes and only the spin-flipping
tunneling is kept.
Otherwise, the spin-conserving and spin-flipping
tunnelings will coexist. Thus the quantum transport with or
without spin-flipping can be manipulated by adjusting the SO
coupling strength $\alpha$. In what follows, we will illustrate how the tunneling dynamics
of a single SO-coupled atom in an optical lattice is governed by interplay between impurity and periodic driving, with focus on the usual spin-conserving tunneling ($\sin\alpha=0$) and
the purely spin-flipping tunneling ($\cos\alpha=0$) case.

\begin{figure}[htb]
\includegraphics[width=0.5\textwidth]{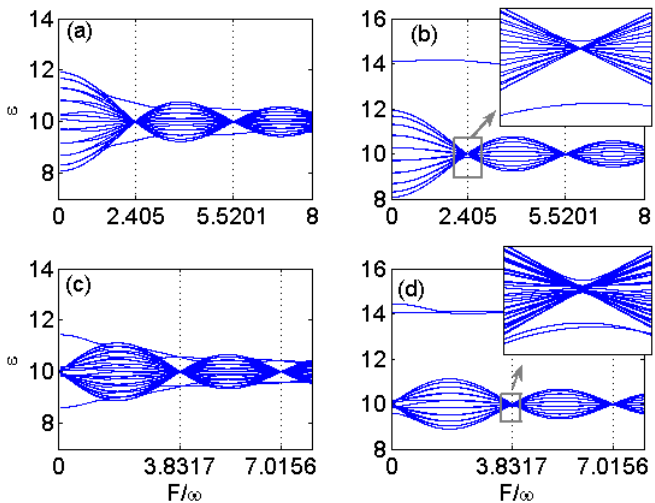}
\caption{(Color online) Numerically computed quasienergy spectrum versus the driving parameter $F/\omega$ for the usual spin-conserving tunneling [$\sin\alpha=0$, upper row (a) and (b)] and
the purely spin-flipping tunneling [$\cos\alpha=0$, lower row (c) and (d)] case. Left [(a) and (c)]: the resonant case $\varepsilon_0/\omega=1$; Right [(b) and (d)]: the nonresonant case $\varepsilon_0/\omega=1.2$. The other parameters are set as $v=1, \omega=20, \Omega=20$. Hereafter, all
variables and parameters are dimensionless. The insets in (b) and (d ) highlight the subtle avoided crossings
 near the collapse points for the nonresonant case.}\label{figure1}
\end{figure}

Before investigating the corresponding quantum tunneling dynamics, we first discuss the quasienergies and Floquet
states of the considered system (\ref{Hamiltonian}). According to Floquet theory, the time-periodicity of Hamiltonian
allows to write solutions of the Schr\"{o}dinger equation (\ref{AmPro}) and (\ref{AmPro2})
in the form of $a_{n,\sigma}(t) = \tilde{a}_{n,\sigma}(t)\exp(-i\varepsilon t)$, where $\varepsilon$ are the quasienergies and $\tilde{a}_{n,\sigma}(t)$,
to which we shall refer as Floquet modes, are periodic with the driving period $\tilde{T}=2\pi/\omega$.

In Fig.~\ref{figure1}, we plot quasienergy spectrum versus the driving parameter $F/\omega$ for the usual spin-conserving tunneling $\sin\alpha=0$ (upper row) and the purely spin-flipping tunneling
$\cos\alpha=0$ (lower row) cases, by direct
diagonalization of the evolution operator over
one period  $U(T, 0) =\mathcal{T} \exp[-i\int_{0}^{\tilde{T}}H(t)dt]$, where $\mathcal{T}$ is the time-ordering operator.
In all the numerical calculations, we have set the size of lattice to be $N=21$ and $\Omega=\omega$.
As  can be seen in the left column of Fig.~\ref{figure1}, in the resonant case of $\varepsilon_0=\omega$ , for $\sin\alpha=0$ (upper left), the quasienergies make up a miniband and exhibit a set of band collapses at $F/\omega=2.405, 5.5201,...$, the zeros of Bessel $J_0$ function, while two pairs of doubly-degenerate quasienergies deviate from the rest of miniband near the collapse regions; however, for $\cos\alpha=0$ (lower left), the quasienergy band shows a similar set of collapses when $F/\omega=3.8317, 7.0156,...$, the zeros of Bessel $J_1$ function, accompanied by four non-degenerate quasienergies  deviating from the rest in the vicinity of the collapse regions.

It is obvious that the quasienergy spectrums corresponding to the case $\sin\alpha=0$, for which only the spin-conserving couplings are presented, have the same structure of
that in the case of their spinless atom counterpart reported in Ref.~\cite{Liang}, where the band collapses have been observed at the zeros of Bessel $J_0$ function. It is because in this situation the results are spin-independent and thus the quasienergy band corresponding  to the spin-up states is exactly the same as that corresponding to the spin-down states.
By contrast, for $\cos\alpha=0$ (only the spin-flipping couplings are presented), the spin-flipping tunneling is associated with an energy cost of $\Omega$, and therefore, when the
energy shift $\Omega$ induced by the spin-flipping tunneling matches a multiple of the driving frequency, that is, when $\Omega=m\omega$ with positive integer $m$, the driven SO-coupled system behaves approximately, in a time-averaged
sense, like
an undriven one with the rescaled hopping matrix element $vJ_m(F/\omega)$. In our case, where $\Omega=\omega$, the collapses of quasienergy band accordingly become located at the zeros of
Bessel $J_1$ function, rather than the zeros of Bessel $J_0$ function as in the spin-independent tunneling case.
Due to the existence of an impurity, there is a corresponding energy difference created by a static tilt between the impurity
and its two nearest-neighbor sites, which results in four associated quasienergies distinguished from quasienergy flatness (narrrowing) in the vicinity of the collapse point, in stark contrast to the case of
perfect optical lattice in which all the quasienergies will collapse into a single value.

As shown in the right column of Fig.~\ref{figure1}, in the off-resonant case of $\varepsilon_0=1.2\omega$,
 except that two
quasienergy levels (which for the usual spin-conserving tunneling $\sin\alpha=0$ case are identical and exactly overlapped, as shown in Fig.~\ref{figure1}(b)) remain isolated from the rest of the miniband for all the values of $F/\omega$, there also exist a series of band collapses just as in the resonant case.
The quasienergy levels separated markedly from the rest correspond to localized states for which an atom with spin up or down
is captured by the impurity.
Particularly interesting, the region of the level collapse
is in fact a region with several avoided crossings as shown in the insets of Figs.~\ref{figure1} (b) and (d).
As we shall show in the next section, these fine structures of avoided level crossings near the pseudocollapse
points can be conveniently exploited for engineering the spin-dependent transportation.

\begin{figure}[htb]
\includegraphics[width=0.5\textwidth]{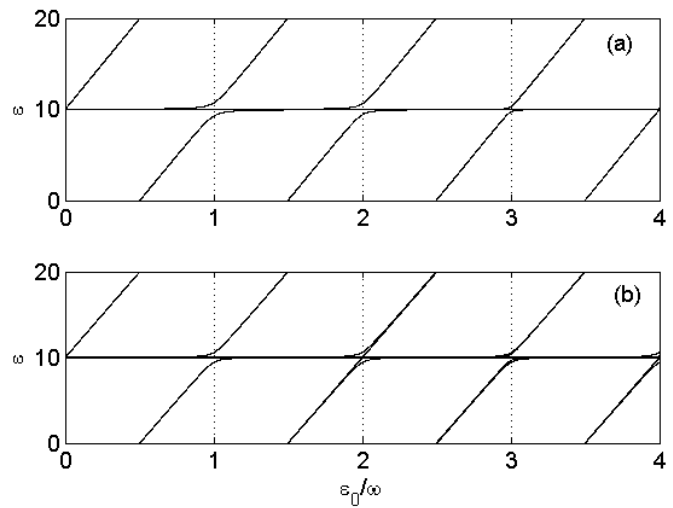}
\caption{One Brillouin zone $0\leq\varepsilon\leq\omega$ of quasienergies as
functions of $\varepsilon_0/\omega$. (a): The usual spin-conserving tunneling ($\sin\alpha=0$) case, $F/w=2.4045$. (b): The purely spin-flipping tunneling ($\cos\alpha=0$) case, $F/w=3.8317$. The other parameters are $v=1, \omega=20, \Omega=20$.}\label{figure2}
\end{figure}

To gain more insight into the behaviors of quasienergies at the collapse points, we plot in Fig.~\ref{figure2} quasienergies within one Brillouin zone $0\leq \varepsilon \leq \omega$
versus $\varepsilon_0/\omega$, by choosing $F/w=2.4045$ [Fig.~\ref{figure2} (a)] for usual spin-conserving tunneling, and $F/w=3.8317$ [Fig.~\ref{figure2} (b)]
for the purely spin-flipping tunneling case respectively.
It is shown in Fig.~\ref{figure2} that close approaches of two branches occur whenever $\varepsilon_0=m'\omega$ with $m'$ being positive integers.
At the points of close approach, the curves will not actually cross, but their separations decrease with increasing integer value of $\varepsilon_0/\omega$.
Other quasienergies are degenerate, representing the quasienergy flatness (collapse) as shown in Fig.~\ref{figure1}.
These close approaches (avoided level crossings) at values of $\varepsilon_0=m'\omega$ indicate the presence of multiple-photon resonances which will be discussed later in detail. The levels which make close approaches (avoided crossings) at values of $\varepsilon_0=m'\omega$, correspond to the four quasienergies differentiating from the rest of the miniband, each two of which will be coincident (or nearly coincident) at the collapse point as shown in Figs.~\ref{figure1} (a) and (c).
However, the fine and subtle structures of avoided level crossings for the off-resonant case depicted in the insets of Figs.~\ref{figure1} (b) and (d)
are too small to be visible in the scale of Fig.~\ref{figure2}.

\begin{figure}[htb]
\includegraphics[width=0.5\textwidth]{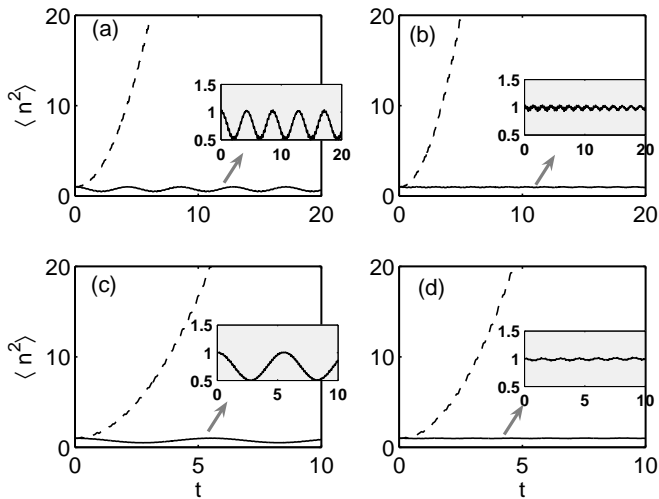}
\caption{Numerically computed time evolution of the
mean-square displacement $\langle n^2 \rangle$, defined by $\langle n^2 \rangle=\sum_{n,\sigma}n^2|a_{n,\sigma}|^2$, for the usual spin-conserving tunneling ($\sin\alpha=0$) case [upper row (a)-(b)] and for the purely spin-flipping tunneling ($\cos\alpha=0$) case [lower row (c)-(d)], respectively. The total number of lattice sites is chosen
as $N=21$. In upper row (a)-(b), the solid lines are for $F/\omega=2.405$ and the dashed lines for $F/\omega=1.5$; while in the lower row (c)-(d), the solid lines are for $F/\omega=3.8317$, and the dashed lines are for $F/\omega=1.5$.  Left column [(a) and (c)]: the resonant case $\varepsilon_0/\omega=1$; Right column [(b) and (d)]: the nonresonant case $\varepsilon_0/\omega=1.2$.  The insets show
 enlargement of the localized dynamics at the first quasienergy band collapse point $F/\omega=2.405$ [upper row (a)-(b)] and $F/\omega=3.8317$  [lower row (c)-(d)]. In all plots, the other parameters are the same as in Fig.~\ref{figure1}, and the system is  initialized
in state $|-1,\uparrow\rangle$, that is, with an initial spin-up atom in the left-side neighbor of impurity site (site $-1$).}\label{figure3}
\end{figure}

\section{tunneling dynamics with or without spin flipping}
In this section, we will present a comprehensive analysis of the tunneling physics underlying the quasienergy
spectrum of the considered system (\ref{Hamiltonian}). As is well known, the crossing (collapse)
of the quasienergies of the underlying time-periodic system is always associated with a celebrated quantum phenomenon called dynamical localization (DL)\cite{Dunlap}, in which a localized particle periodically
returns to its initial state following the periodic change of
the field. Upon occurrence of DL, the initial localized quantum state
will not diffuse. To study the system's time-evolution
quantitatively, we have plotted the time dependence of the
mean-square displacement $\langle n^2 \rangle$ ($=\sum_{n,\sigma}n^2|a_{n,\sigma}|^2$), as illustrated in Fig.~\ref{figure3}, by numerically solving equations (\ref{AmPro}) and (\ref{AmPro2}) with the size of lattice $N=21$.
Clearly,  for the off-resonance case, if we start the system with a single spin-up (or spin-down) atom in the impurity site (site $n=0$), the system  will be frozen in the initial state, which is suggested by the structure of two levels isolated from the rest of miniband in Figs.~\ref{figure1} (b) and (d).
The frozen dynamics arises from the fact the particle is captured by impurity.
Thus, in order to illustrate the physical significance of the miniband collapses, we assume in the following calculations that the system be initialized
in state $|-1,\uparrow\rangle$, that is, with a spin-up atom in the left-side neighbor of impurity site (site $-1$).
For this initial state, if there is a diffusive (spatially delocalized) propagation  through the whole optical lattice, the
mean-square displacement $\langle n^2 \rangle$  will increase without bound.
By comparison of Figs.~\ref{figure3} (a)-(b) and (c)-(d), it can be readily seen that the behaviors of time-dependence of
the mean-square displacement are similar between the usual spin-conserving tunneling $\sin\alpha=0$ (top) and
the purely spin-flipping tunneling $\cos\alpha=0$ (bottom) case, whether the driving parameters are located on (the solid lines) or off (the dashed lines) the band collapse points.
As is apparent in Fig.~\ref{figure3}, for both $F/\omega=2.405$ (solid lines in the top row) and $F/\omega=3.8317$ (solid lines in the bottom row) corresponding to the collapse points in the cases of $\sin\alpha=0$ and
$\cos\alpha=0$ respectively, the mean-square displacements
are seen to be bounded and the celebrated dynamical localization (DL)\cite{Dunlap} occurs, whereas in the case of $F/\omega=1.5$ (dashed lines)
the mean-square displacements grow without bound and the particle initially occupying site -1 delocalizes.
Detailed examination of the
bounded solutions (evidence of DL) reveals that the mean-square displacement varies between $1$ and $0.5$ for the resonant case (see insets in left column), indicating that the particle propagates through several lattice spacings and returns repeatedly to the initial location,
while the mean-square displacement remains near unity for the off-resonant case (see insets in right column), seeming to be a frozen dynamics.
The seemingly frozen dynamics, as we will show below, is actually a second-order tunneling between the two nearest-neighbor sites of the impurity.

\begin{figure}[htb]
\includegraphics[width=0.5\textwidth]{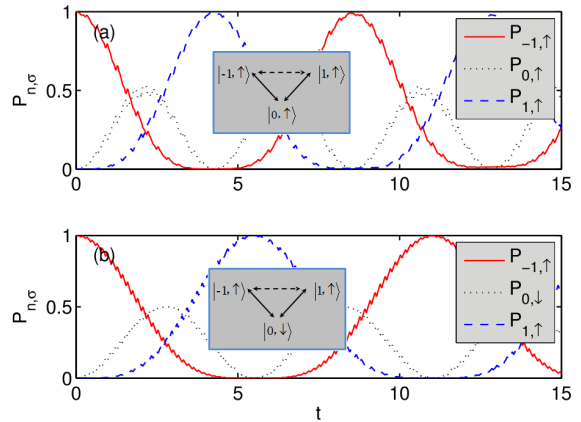}
\caption{(Color online) Time evolution of occupation probabilities $P_{n,\sigma¦Ò}$ given by equations (\ref{AmPro}) and (\ref{AmPro2}) for the resonant case $\varepsilon_0/\omega=1$, starting the system with a spin-up particle in site $n=-1$. (a): The usual spin-conserving tunneling ($\sin\alpha=0$) case at the first collapse point $F/\omega=2.405$. (b): The purely spin-flipping tunneling ($\cos\alpha=0$) case at the first collapse point $F/\omega=3.8317$. The total number of lattice sites is set as $N=21$, and other parameters are $v=1, \omega=20, \Omega=20$.}\label{figure4}
\end{figure}

 For a clearer observation of the tunneling dynamics presented above, we show in Fig.~\ref{figure4} the time evolution of occupation probabilities  $P_{n ,\sigma}$ for the resonant case $\varepsilon_0/\omega=1$ via direct integration of equations (\ref{AmPro}) and (\ref{AmPro2}) with the size of lattice $N=21$, assuming that
a spin-up atom is initially prepared in the site $-1$ and that the driving parameters are properly set at the band collapse
points. Under such assumptions, though the dynamical localization dominates globally, resonant oscillations with (without) spin flipping can take place
between the impurity and its two nearest-neighbor sites, as depicted in  Fig.~\ref{figure4} (a) and (b) respectively.  For the same set of parameters
as in inset of Fig.~\ref{figure3} (a), where $\sin\alpha=0$ and  $F/\omega=2.405$, the occupation probabilities $P_{-1 ,\uparrow}$ ($P_{1 ,\uparrow}$) oscillate between $1$ and $-1$, at the same time, the occupation probability $P_{0,\uparrow}$ varies between $0$ and $0.5$, which indicates that the dynamics of system is dominated by the spin-conserving tunneling only between the impurity and its two nearest neighbor sites and thus can be understood in an effective three-site model with three localized basis functions $|0,\uparrow\rangle$ and $|\pm 1,\uparrow\rangle$, see Fig.~\ref{figure4} (a) and its inset.
Likewise,  for the same
values of the parameters as in inset of Fig.~\ref{figure3} (c), where $\cos\alpha=0$ and $F/\omega=3.8317$, a complete oscillation between the sites $0$ and $\pm 1$, same as in Fig.~\ref{figure4} (a) but with spin flipping and longer oscillation period, is observed in Fig.~\ref{figure4} (b), in which case the system
dynamics is limited in a subspace spanned by states $|0,\downarrow\rangle$ and $|\pm 1,\uparrow\rangle$ and thus can be described in an
effective three-site model [the inset in Fig.~\ref{figure4} (b)]. These resonant oscillations (shown in Figs.~\ref{figure4} (a) and (b)) between the impurity and its two nearest-neighbor sites are strongly reminiscent of multiple-photon resonances, which manifest themselves in the avoided crossings of quasienergy spectrum at the band collapse points (shown in Figs.~\ref{figure1} (a) and (c)) and under the circumstances the system will exchange energy of an integer number of photons with the
oscillating field to overcome the energy offset created by the impurity, thus offering a possible means for control of single-spin transportation with or without spin flipping. Noticing that the results are physically similar if we instead start the system with a single spin-down atom in the site $-1$, without loss of generality we restrict ourselves to the case of initial spin-up atom throughout our paper.

\begin{figure}[htb]
\includegraphics[width=0.5\textwidth]{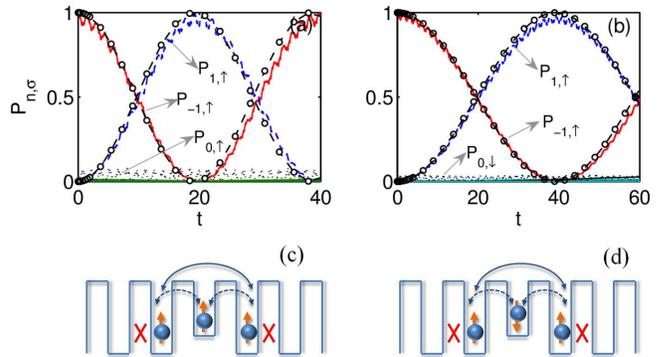}
\caption{(Color online) Upper row (a) and (b): Time evolution of occupation probabilities $P_{n,\sigma¦Ò}$ given by equations (\ref{AmPro}) and (\ref{AmPro2})
(with the size of lattice $N=21$) for the nonresonant case $\varepsilon_0/\omega=1.2$. (a): The usual spin-conserving tunneling ($\sin\alpha=0$) case at the first collapse point $F/\omega=2.405$. (b): The purely spin-flipping tunneling ($\cos\alpha=0$) case at the first collapse point $F/\omega=3.8317$.
The circles in (a) and (b), respectively, represent the second-order perturbative results from equation (\ref{order1}) and (\ref{order2}).
Lower panels in (c) and (d) are schematic representations of the tunneling
dynamics presented in (a) and (b) respectively. Red crosses indicate suppression of tunneling through that barrier, and the dashed-line arrows indicate the virtual first-order tunneling process, whereas the
solid-line arrows indicate the
allowed second-order tunneling which actually occurs. The other system parameters and initial state are the same as in Fig.~\ref{figure4}.}\label{figure5}
\end{figure}

Now we turn to the off-resonant case with the same set of parameters and initial conditions
as in insets of Figs.~\ref{figure3} (b) and (d) where the dynamical localization takes place globally. In this case,
the transition between the impurity and its two nearest-neighbor sites is
nonresonant and is therefore  suppressed when the strength of the impurity potential is comparatively
far from any integer
multiple of $\omega$. To our surprise, the considered system actually performs
Rabi oscillation between the localized states $|-1,\uparrow\rangle$ and $|1,\uparrow\rangle$ with negligible
population at all the remaining sites ($n\neq\pm 1$) for both the  usual spin-conserving tunneling [Fig.~\ref{figure5} (a)] and
the purely spin-flipping tunneling [Fig.~\ref{figure5} (b)]  cases.
These two Rabi oscillations shown in Figs.~\ref{figure5} (a) and (b) correspond to the very similar observable tunneling behaviors where a single spin-up atom tunnels back and forth between the two nearest-neighbor sites of
the impurity in a manner that the spin ($\uparrow$) remains unchanged and the impurity site population is negligible over all the evolution time, and thus their corresponding mean-square displacements as
illustrated in the insets of Figs.~\ref{figure3} (b) and (d) apparently will
keep  unchanged because of the symmetry of lattice.
Obviously, the tight-binding model (\ref{Hamiltonian}) can only give rise to the first-order transition between two adjacent sites,
so that these tunneling processes between the two sites $-1$ and $1$ are possible only due to the second-order transition.
As we shall show in the next section in detail, two distinct
physical mechanisms are believed to be involved. The first one is due to the second-order transition $|-1,\uparrow\rangle\rightarrow |1,\uparrow\rangle$
via the virtual intermediate state $|0,\uparrow\rangle$. The second one is induced by the second-order transition $|-1,\uparrow\rangle\rightarrow |1,\uparrow\rangle$ via
a different virtual intermediate state $|0,\downarrow\rangle$.
The respective
tunneling dynamics can be summarized schematically in lower panels
in Figs.~\ref{figure5} (c) and (d), where crosses
indicate suppression of tunneling
through that barrier, the dashed-line arrows indicate the virtual first-order tunneling process,
whereas the solid-line arrows indicate the second-order tunneling
allowed.

In a nutshell, there are two types of second-order tunneling processes
despite of very similar observable tunneling behaviors, which are interpreted
in the context of two different mechanisms as schematically shown in Figs.~\ref{figure5} (c) and (d).
Actually, the aforementioned second-order tunneling processes are the direct consequences of the subtle avoided level crossings presented in the insets
of Figs.~\ref{figure1} (b) and (d), which can not be captured by the first-order perturbation and needs to be explained in
the framework of the second-order perturbative theory.

\section{Analysis of the tunneling dynamics}

According to the above numerical results, we preliminarily  come to a conclusion that the dynamical localization of the two-component atomic gas with SO coupling is non-destroyed by the impurity for both the usual spin-conserving tunneling and
the purely spin-flipping tunneling cases, however, the local dynamics of the system are dramatically tuned by the ratio of the impurity potential to the driving frequency. Under the dynamical localization (DL) condition, in other words, when the driving parameters are properly chosen at the band collapse points, the occurrences of resonant tunneling as well as the second-order tunneling have been observed, in which the single atom can move only among the three sites $0$ and $\pm 1$ while all other inter-site tunneling passages are shut off. In both cases, the tunneling dynamics can be understood by an effective three-site model and the time evolution of quantum state of a spin-$1/2$ particle should be confined in the truncated Hilbert spaces spanned by six Fock states. Next, we will try to substantiate our numerical calculations by the analytical method.

To proceed, we begin with Eqs.~(\ref{AmPro}) and (\ref{AmPro2}) and introduce the following new amplitudes:
\begin{eqnarray}
 \label{subst}
  b_{n,\uparrow}(t)&=&{a}_{n,\uparrow}(t)\exp\{i[n\Phi(t)+\frac{\Omega}{2}t+\varepsilon_{0}\delta_{n,0}t]\},  \label{subst} \\
  b_{n,\downarrow}(t)&=&{a}_{n,\downarrow}(t)\exp\{i[n\Phi(t)-\frac{\Omega}{2}t+\varepsilon_{0}\delta_{n,0}t]\}, \label{subst2}
\end{eqnarray}
where we have set $\Phi(t)=\int_0^{t} d\tau\varepsilon'(\tau)=\frac{F}{\omega}\sin\omega t$.

In terms of the new amplitudes $ b_{n,\sigma} (\sigma=\uparrow,\downarrow)$, we can rewrite equations (\ref{AmPro}) and (\ref{AmPro2}) in the form

\begin{align}
     i\dot{b}_{n,\uparrow}=&-\upsilon\cos\alpha[{b}_{n+1,\uparrow}e^{-i\Phi(t)}+{b}_{n-1,\uparrow}e^{i\Phi(t)}]\nonumber\\
     &-\upsilon\sin\alpha[{b}_{n-1,\downarrow}e^{i\Phi(t)+i\Omega t}-{b}_{n+1,\downarrow}e^{-i\Phi(t)+i\Omega t}],\nonumber\\
     &(n\neq-1 ,0,1),\nonumber\\
     i\dot{b}_{n,\downarrow}=&-\upsilon\cos\alpha[{b}_{n+1,\downarrow}e^{-i\Phi(t)}+{b}_{n-1,\downarrow}e^{i\Phi(t)}]\nonumber\\
     &-\upsilon\sin\alpha[{b}_{n+1,\uparrow}e^{-i\Phi(t)-i\Omega t}-{b}_{n-1,\uparrow}e^{i\Phi(t)-i\Omega t}],\nonumber\\
     &(n\neq-1 ,0,1),\nonumber\\
     i\dot{b}_{-1,\uparrow}=&-\upsilon\cos\alpha[{b}_{0,\uparrow}e^{-i\Phi(t)-i\varepsilon_{0}t}+{b}_{-2,\uparrow}e^{i\Phi(t)}]\nonumber\\
     &-\upsilon\sin\alpha[{b}_{-2,\downarrow}e^{i\Phi(t)+i\Omega t}-{b}_{0,\downarrow}e^{-i\Phi(t)+i(\Omega-\varepsilon_{0}) t}],\nonumber\\
     i\dot{b}_{-1,\downarrow}=&-\upsilon\cos\alpha[{b}_{0,\downarrow}e^{-i\Phi(t)-i\varepsilon_{0}t}+{b}_{-2,\downarrow}e^{i\Phi(t)}]\nonumber\\
     &-\upsilon\sin\alpha[{b}_{0,\uparrow}e^{-i\Phi(t)-i(\Omega+\varepsilon_{0}) t}-{b}_{-2,\uparrow}e^{i\Phi(t)-i\Omega t}],\nonumber\\
     i\dot{b}_{0,\uparrow}=&-\upsilon\cos\alpha[{b}_{1,\uparrow}e^{-i\Phi(t)+i\varepsilon_{0}t}+{b}_{-1,\uparrow}e^{i\Phi(t)+i\varepsilon_{0}t}]\nonumber\\
     &-\upsilon\sin\alpha[{b}_{-1,\downarrow}e^{i\Phi(t)+i(\Omega+\varepsilon_{0}) t}-{b}_{1,\downarrow}e^{-i\Phi(t)+i(\Omega+\varepsilon_{0}) t}],\nonumber\\
     i\dot{b}_{0,\downarrow}=&-\upsilon\cos\alpha[{b}_{1,\downarrow}e^{-i\Phi(t)+i\varepsilon_{0}t}+{b}_{-1,\downarrow}e^{i\Phi(t)+i\varepsilon_{0}t}]\nonumber\\
     &-\upsilon\sin\alpha[{b}_{1,\uparrow}e^{-i\Phi(t)-i(\Omega-\varepsilon_{0}) t}-{b}_{-1,\uparrow}e^{i\Phi(t)-i(\Omega-\varepsilon_{0}) t}],\nonumber\\
      i\dot{b}_{1,\uparrow}=&-\upsilon\cos\alpha[{b}_{2,\uparrow}e^{-i\Phi(t)}+{b}_{0,\uparrow}e^{i\Phi(t)-i\varepsilon_{0}t}]\nonumber\\
     &-\upsilon\sin\alpha[{b}_{0,\downarrow}e^{i\Phi(t)+i(\Omega-\varepsilon_{0}) t}-{b}_{2,\downarrow}e^{-i\Phi(t)+i\Omega t}],\nonumber\\
     i\dot{b}_{1,\downarrow}=&-\upsilon\cos\alpha[{b}_{2,\downarrow}e^{-i\Phi(t)}+{b}_{0,\downarrow}e^{i\Phi(t)-i\varepsilon_{0}t}]\nonumber\\
     &-\upsilon\sin\alpha[{b}_{2,\uparrow}e^{-i\Phi(t)-i\Omega t}-{b}_{0,\uparrow}e^{i\Phi(t)-i(\Omega+\varepsilon_{0}) t}].
     \nonumber\\
     \label{NAmPro}
    \end{align}
Given the on-site energy at site $n=0$ modified by the impurity, as can be seen in Eq.~(\ref{NAmPro}), the equations of motion describing the system with a spin-$1/2$ particle in
sites $n=0, \pm 1$ are  in form different from those in other sites,
which  will lead to a remarkable change in the local dynamics of  the sites $n=0, \pm 1$.
For obtaining a close correspondence in analytic and numerical results, here we typically assume that the model (\ref{Hamiltonian}) is in high-frequency ($\omega\gg v$) limits and $\Omega=m\omega,\varepsilon_{0}=m'\omega+u$ for $|u|\leq \omega/2$, $m, m'=1,2,...$ with $u$ being the reduced impurity strength.

By means of high-frequency approximation method, in which the rapidly oscillating exponential functions in the equation (\ref{NAmPro}) are replaced by
their time average and the formula $\exp \left[ {\ \pm
iF\sin \left( {\omega t} \right)/\omega } \right] = \sum\nolimits_k {J_k
\left( {F/\omega } \right)\exp \left( {\ \pm ik\omega t} \right)}$ is utilized,
we easily figure out that the hopping amplitude $v$ between the states with the same (different) spins and across the neighboring sites $i$ and $j$ ($i,j\neq 0$) is
renormalized by $vJ_0(F/\omega)$ and $vJ_m(F/\omega)$ respectively.
Nevertheless, the effective hopping amplitudes between the neighboring sites $i$ and $j$ (either $i=0$ or $j=0$)
are determined by the reduced impurity strength as well as the ratio of the impurity potential to the driving frequency.
Thus, the dynamical localization (DL), also referred to as
 coherent destruction of tunneling (CDT) in the high-frequency limit\cite{Grossmann}, occurs at distinct values of the scaled driving amplitude $F/\omega$, which satisfies
either $J_0(F/\omega)=0$ for the usual spin-conserving tunneling ($\sin\alpha=0$) case or $J_m(F/\omega)=0$ for
the purely spin-flipping tunneling ($\cos\alpha=0$) case.
When DL (CDT) occurs, the dynamics of sites $0,\pm 1$ is decoupled from that of other sites, and only tunneling among sites $0,\pm 1$ is allowed.
\subsection{Tunneling dynamics with only spin-conserving coupling}
In this subsection, we will analyze the case of the usual spin-conserving tunneling ($\sin\alpha=0$), where the DL condition is given by  $J_0(F/\omega)=0$.
Under the DL condition $J_0(F/\omega)=0$, the system dynamics of spin-$1/2$ particle will be limited in two independent subspaces: $\{|-1, \uparrow\rangle, |0,\uparrow\rangle, |1,\uparrow\rangle\}$ and $\{|-1, \downarrow\rangle, |0,\downarrow\rangle, |1,\downarrow\rangle\}$.
In this case, however, the tunneling dynamics among sites $0,\pm 1$ are partially restored and tuned by the
value of the impurity potential relative to the driving frequency.

Let us first consider the resonant case, namely, $\varepsilon_{0}=m'\omega, m'=1,2,...$. By averaging the rapidly oscillating exponential terms in equations of $b_{n,\sigma}(t) (n=0,\pm 1)$, we have the effective three-site model in the truncated subspaces
\begin{align}\label{NAmProRes2}
      i\dot{b}_{-1,\sigma}=&-\upsilon\cos\alpha J_{-m'}(\frac{F}{\omega}){b}_{0,\sigma},\nonumber\\
     i\dot{b}_{0,\sigma}=&-\upsilon\cos\alpha[ J_{m'}(\frac{F}{\omega}){b}_{1,\sigma}+ J_{-m'}(\frac{F}{\omega}){b}_{-1,\sigma}],\nonumber\\
     i\dot{b}_{1,\sigma}=&-\upsilon\cos\alpha J_{m'}(\frac{F}{\omega}){b}_{0,\sigma},
\end{align}
with $\cos\alpha =\pm 1$. As can be seen in equation (\ref{NAmProRes2}), the hopping matrix for nearest neighbors is spin independent and the dynamics of the system in the subspace $\{|-1 ,\downarrow\rangle,|0 ,\downarrow\rangle,|1 ,\downarrow\rangle\}$ is exactly the same as in the subspace $\{|-1 ,\uparrow\rangle,|0 ,\uparrow\rangle,|1 ,\uparrow\rangle\}$.
It explains why in this case the quasienergies corresponding to the Floquet states with spin up are exactly overlapped with those with spin down.
The combined actions of the impurity and the high-frequency periodic driving are expected to create Floquet-quasienergy
spectrum, in which the hopping matrix between the impurity and any one of its nearest neighbors is renormalized to an effective hopping parameter $vJ_{\pm m'}(F/\omega)$ rather than to $vJ_{0}(F/\omega)$ such that avoided  level crossings appear at the band collapse points $J_{0}(F/\omega)=0$ as shown in Fig.~\ref{figure1} (a).  In the resonant case, the system dynamics is dominated by the resonant tunneling between the impurity and its two nearest-neighbor sites as shown in Fig.~\ref{figure4} (a) despite in the presence of strong tilt, which is the familiar $m'$-photon assisted tunneling for $m'=1,2,...$..

It should be born in mind that the commonly used high-frequency (averaging) approximation, which has also been exploited for derivation of the above equation (\ref{NAmProRes2}), corresponds to the first-order perturbation approximation.
In this perturbation approximation treatment, for relatively strong values of the reduced impurity strength $u$, the oscillation of the functions $e^{\pm iut}$ in equation (\ref{NAmPro}) becomes moderately fast and should be replaced by their average value (zero) such that
 the first-order transitions between sites $0$ and $-1$ (or $1$) are frozen approximately. Obviously, this remarkable second-order tunneling effect on the basis of
the numerical result is missed in the high-frequency approximation  analysis. A more
accurate perturbative analysis is therefore needed for the correct study of the dynamical
behaviors.

Now we continue with the off-resonant regime, $\varepsilon_{0}=m'\omega+u, m'=1,2,...$ with moderate value of $|u|$, so as to investigate the tunneling
dynamics of the system with only spin-conserving coupling at the collapse point $J_0(F/\omega)=0$, by means of the multiple-time-scale asymptotic analysis\cite{Longhi2,Zhou2,Luo}.
Equipped with the effective three-site model in a truncated Hilbert space spanned by the Fock basis set  $\{|-1,\sigma\rangle, |0,\sigma\rangle, |1,\sigma\rangle\}$ ($\sigma=\uparrow,\downarrow$), we start by introducing $\epsilon=\upsilon/\omega$ as a small
parameter for high-frequency driving and the normalized time
variable $\tau=\omega t$, and by writing $b_{n,\sigma}$ ($n=0,\pm 1, \sigma=\uparrow,\downarrow$) as a power-series expansion in $\epsilon$,
\begin{equation}\label{powerseries}
    b_{n,\sigma}(\tau)=b_{n,\sigma}^{(0)}(\tau)+\epsilon b_{n,\sigma}^{(1)}(\tau)+\epsilon^{2} b_{n,\sigma}^{(2)}(\tau)+\cdot\cdot\cdot.
\end{equation}
 Because the high-order terms can be neglected in the high-frequency regime, the probability amplitudes $b_{n,\sigma}$
can be approximately rewritten as the zeroth order $ b_{n,\sigma}(\tau)=b_{n,\sigma}^{(0)}(\tau)=A_{n,\sigma}(\tau)$. By definition, $|A_{n,\sigma}|^2$ ($n=0,\pm 1$) refers to the occupation probability for a single particle with spin $\sigma$ in site $n$.
According to the perturbation analysis in the appendix A, we  obtain the following equations for the amplitudes $A_{n,\sigma}$ ($n=0,\pm 1$) up to the second-order $\epsilon^{2}$ for the usual spin-conserving tunneling ($\sin\alpha=0$) case
\begin{align}\label{order1}
    i\frac{dA_{-1,\sigma}}{d\tau}&=-\epsilon^2(\chi_{1}A_{1,\sigma}+\chi_{2}A_{-1,\sigma})\nonumber
    \\i\frac{dA_{0,\sigma}}{d\tau}&=2\epsilon^2\chi_{2}A_{0,\sigma}\nonumber
    \\i\frac{dA_{1,\sigma}}{d\tau}&=-\epsilon^2(\chi_{2}A_{1,\sigma}+\chi_{1}A_{-1,\sigma}),
\end{align}
where we have set
\begin{align}
  \chi_{1} &=\sum\limits_{p}\frac{J_{p}(\frac{F}{\omega})J_{-p}(\frac{F}{\omega})}{-p+m'+u'},~~~~
    \chi_{2}&=\sum\limits_{p}\frac{J_{p}^{2}(\frac{F}{\omega})}{p+m'+u'}.
\end{align}

Equation (\ref{order1}) provides a correct description of the dynamics of the original system with only spin-conserving coupling
under DL condition up to the second-order long time scale $\sim 1/\epsilon^2$. In deriving Eq.~(\ref{order1}), the intermediate state
$|0,\sigma \rangle$ is eliminated and accordingly an effective tunneling rate between states $|-1,\sigma\rangle$ and $|1,\sigma\rangle$ is
given by $\omega\epsilon^2\chi_{1}\equiv v^2\chi_{1}/\omega\ll v$, through returning to the original time variable $t$.
From equation (\ref{order1}), we immediately note that the dynamics of the impurity is decoupled from that of its two neighbors and therefore the particle will be
captured by the impurity if the system is initially prepared with a single particle in the impurity site.
Considering that the temporal evolution equations for amplitude $A_{n,\sigma}$ ($n=0,\pm 1$) are the same for any spin, as can be seen from equation (\ref{order1}), we thus
have the spin-independent second-order tunneling between the two nearest-neighbor sites of the impurity with zero population at the impurity site.
In Fig.~\ref{figure5}(a), we have calculated the time evolution of occupation probabilities $P_{n,\sigma}$
via integration of equation (\ref{order1})--the second-order perturbative results, and compared them (circles) with the numerical results obtained from the
original model (\ref{Hamiltonian}) for the usual spin-conserving tunneling ($\sin\alpha=0$) case with the initial state $|\psi(t=0)\rangle=|-1,\uparrow\rangle$. A good agreement is found.

According to the Floquet theorem, we can construct the analytical Floquet solutions of the system by setting  $a_{n,\sigma}=\tilde{a}_{n,\sigma}(t)\exp(-i\varepsilon t)=A_{n,\sigma}(t)\exp\{-i[n\Phi(t)+(-1)^k\frac{\Omega}{2}t+\varepsilon_{0}\delta_{n,0}t]\}=A'_{n,\sigma}\exp\{-i[n\Phi(t)+(-1)^k\frac{\Omega}{2}t+\varepsilon_{0}\delta_{n,0}t]\}\exp(-iEt)$ ($k=0,1$ for $\sigma=\uparrow,\downarrow$, respectively) with the help of Eqs.~(\ref{subst}) and (\ref{subst2}), where constants $A'_{n,\sigma}$ and $E$ are the eigenvector components and the eigenvalue of the time-independent version of equation (\ref{order1}) respectively. By inserting $A_{n,\sigma}(t)=A'_{n,\sigma}\exp(-iEt)$ into Eq.~(\ref{order1}), and employing the transformations (\ref{subst}) and (\ref{subst2}), we obtain the quasienergies for the off-resonant case with moderate value of $u$
\begin{align}
\label{Qeor1}
\varepsilon_{1,\sigma}&=\frac{-v^2\chi_2+v^2\chi_1}{\omega}+(-1)^k\frac{\Omega}{2},\nonumber\\
    \varepsilon_{2,\sigma}&=\frac{2v^2\chi_2}{\omega}+u+(-1)^k\frac{\Omega}{2},\nonumber\\
    \varepsilon_{3,\sigma}&=\frac{-v^2\chi_2-v^2\chi_1}{\omega}+(-1)^k\frac{\Omega}{2},
\end{align}
with the corresponding Floquet modes
\begin{align}
|\varepsilon_{1,\sigma}(t)\rangle&=\large(\frac{1}{\sqrt{2}}\exp[i\Phi(t)],0,-\frac{1}{\sqrt{2}}\exp[-i\Phi(t)]\large)^{T},\nonumber\\
    |\varepsilon_{2,\sigma}(t)\rangle&=\large(0,\exp(im'\omega t),0\large)^{T},\nonumber\\
    |\varepsilon_{3,\sigma}(t)\rangle&=\large(\frac{1}{\sqrt{2}}\exp[i\Phi(t)],0,\frac{1}{\sqrt{2}}\exp[-i\Phi(t)]\large)^{T},
    \label{Fmode}
\end{align}
where we have decomposed $\varepsilon_{0}$ into a form of  $\varepsilon_{0}=m'\omega+u, m'=1,2,\cdot\cdot\cdot$, and formulated the Floquet modes  $|\varepsilon_{j,\sigma}(t)\rangle\equiv (\tilde{a}_{-1,\sigma}, \tilde{a}_{0,\sigma}, \tilde{a}_{1,\sigma})^{T}$ for $j=1,2,3$.
Note that the functions of $\exp[\pm i\Phi(t)]$ and $\exp(im'\omega t)$ are $T$-periodic and hence that the Floquet modes inherit the period of the driving.
The two sets of quasienergies ($k=0,1$) are for different spins $\sigma=\uparrow,\downarrow$ respectively, which are almost equal but displaced by a constant amount $\Omega$. Due to the fact that the quasienergies possess Brillouin zonelike structure with one zone width  being $\omega$, when $\Omega=m\omega, m=1,2,\cdot\cdot\cdot$, the two
sets of quasienergies can be mapped to one Brillouin
zone and become identical if one set of quasienergies are all shifted by a constant amount $\Omega$.
In this case, when falling into one Brillouin zone, $0<\varepsilon\leq\omega$, the two sets of quasienergies for different spins are equal to each other and given by
\begin{align}
\label{Qeor2}
\varepsilon_{1,\sigma}&=\frac{-v^2\chi_2+v^2\chi_1}{\omega}+\frac{\omega}{2},\nonumber\\
    \varepsilon_{2,\sigma}&=\frac{2v^2\chi_2}{\omega}+u+\frac{\omega}{2},\nonumber\\
    \varepsilon_{3,\sigma}&=\frac{-v^2\chi_2-v^2\chi_1}{\omega}+\frac{\omega}{2}.
\end{align}
\subsection{Tunneling dynamics with only spin-flipping coupling}
Now we  are in a position to consider the case of the purely spin-flipping tunneling ($\cos\alpha=0$), where the DL condition is given by  $J_m(F/\omega)=0$ when $\Omega=m\omega$ with $m$ positive integers. In this case, when the DL condition $J_m(F/\omega)=0$ is satisfied, the system dynamics of spin-$1/2$ particle will be limited in the following two independent subspaces: $\{|-1,\uparrow\rangle, |0,\downarrow\rangle, |1,\uparrow\rangle\}$ and $\{|-1,\downarrow\rangle, |0,\uparrow\rangle, |1,\downarrow\rangle\}$. Next, we will perform qualitative analysis on the resonant tunneling and
second-order tunneling for the system with only spin-flipping coupling under DL condition, as follows.

When the impurity potential strength is in resonance with the driving field, $\varepsilon_{0}=m'\omega, m'=1,2,...$, the high-frequency  averaging approximation method yields the temporal evolution equation governing the local dynamics of the three sites $0,\pm 1$,
\begin{align}\label{NAmProRes1}
    i\dot{b}_{-1,\uparrow}=&\upsilon\sin\alpha J_{m-m'}(\frac{F}{\omega}){b}_{0,\downarrow},\nonumber\\
     i\dot{b}_{0,\downarrow}=&-\upsilon\sin\alpha[J_{-m+m'}(\frac{F}{\omega}){b}_{1,\uparrow} -J_{m-m'}(\frac{F}{\omega}){b}_{-1,\uparrow}],\nonumber\\
      i\dot{b}_{1,\uparrow}=&-\upsilon\sin\alpha J_{-m+m'}(\frac{F}{\omega}){b}_{0,\downarrow},\nonumber\\
      i\dot{b}_{-1,\downarrow}=&-\upsilon\sin\alpha J_{-m-m'}(\frac{F}{\omega}){b}_{0,\uparrow} ,\nonumber\\
     i\dot{b}_{0,\uparrow}=&-\upsilon\sin\alpha[-J_{m+m'}(\frac{F}{\omega}){b}_{1,\downarrow}
     +J_{-m-m'}(\frac{F}{\omega}){b}_{-1,\downarrow}],\nonumber\\
     i\dot{b}_{1,\downarrow}=&\upsilon\sin\alpha J_{m+m'}(\frac{F}{\omega}){b}_{0,\uparrow},
\end{align}
with $\sin\alpha=\pm 1$. We again encounter the familiar multiphoton
resonances, for which the system is able to exchange energy of an integer number of photons with the
oscillating field to bridge the energy difference created by both the impurity potential and the spin-flipping tunneling due to SOC,
and thus the tunneling is partly restored.  This is witnessed by the fact we have nonzero renormalized tunneling rate ($J_{m-m'}(F/\omega)\neq 0$ and $J_{m+m'}(F/\omega)\neq 0$ in Eq.~(\ref{NAmProRes1})) even at the collapse points corresponding to the zeros of $J_m(F/\omega)$.
As the effective hopping rates are distinct when the dynamics is confined in  the two independent subspaces: $\{|-1,\uparrow\rangle, |0,\downarrow\rangle, |1,\uparrow\rangle\}$ and $\{|-1,\downarrow\rangle, |0,\uparrow\rangle, |1,\downarrow\rangle\}$ respectively, Eq.~(\ref{NAmProRes1}) describes the spin-dependent resonant tunneling with spin flipping among the three sites $0,\pm 1$, which is consistent with the numerical results presented in Fig.~\ref{figure4} where we have set $m=m'=1$ and the initial state $|-1,\uparrow\rangle$.

Turning to the off-resonant case, $\varepsilon_{0}=m'\omega+u, m'=1,2,...$, with moderate value of $|u|$, where the second-order tunneling between the two nearest-neighbor sites of the impurity  will emerge even under the DL condition.  In the
truncated Hilbert space, we still proceed with expanding the amplitudes $b_{n,\sigma}$ ($n=0,\pm 1, \sigma=\uparrow,\downarrow$) as a power-series of $\epsilon$,
$ b_{n,\sigma}(\tau)=b_{n,\sigma}^{(0)}(\tau)+\epsilon b_{n,\sigma}^{(1)}(\tau)+\epsilon^{2} b_{n,\sigma}^{(2)}(\tau)+\cdot\cdot\cdot$, where $\tau=\omega t, \epsilon=v/\omega$, and approximate the probability amplitudes $b_{n,\sigma}$
as the leading order $ b_{n,\sigma}(\tau)=b_{n,\sigma}^{(0)}(\tau)=A_{n,\sigma}(\tau)$.
It should be noted that the other amplitudes $b_{n,\sigma}$ ($n\neq 0,\pm 1$) do not change in time, due to the frozen dynamics of lattice sites for $n\neq 0,\pm 1$ under the DL condition.
According to the standard multiple-time-scale asymptotic analysis method (see the details in the Appendix B), we  obtain the evolution equations for the amplitudes $A_{n,\sigma}$ ($n=0,\pm 1$) up to the second-order $\epsilon^{2}$ for the purely spin-flipping tunneling ($\cos\alpha=0$) case
\begin{align}\label{order2}
    i\frac{dA_{-1,\uparrow}}{d\tau}&=\epsilon^{2}(\chi_{3}A_{1,\uparrow}-\chi_{4}A_{-1,\uparrow})\nonumber
    \\i\frac{dA_{0,\downarrow}}{d\tau}&=2\epsilon^{2}\chi_{4}A_{0,\downarrow}\nonumber
    \\i\frac{dA_{1,\uparrow}}{d\tau}&=-\epsilon^{2}(\chi_{4}A_{1,\uparrow}-\chi_{3}A_{-1,\uparrow})\nonumber
     \\i\frac{dA_{-1,\downarrow}}{d\tau}&=\epsilon^{2}(\chi_{5}A_{1,\downarrow}-\chi_{6}A_{-1,\downarrow})\nonumber
    \\i\frac{dA_{0,\uparrow}}{d\tau}&=2\epsilon^{2}\chi_{6}A_{0,\uparrow}\nonumber
    \\i\frac{dA_{1,\downarrow}}{d\tau}&=-\epsilon^{2}(\chi_{6}A_{1,\downarrow}-\chi_{5}A_{-1,\downarrow}),
\end{align}
where we have set
\begin{align}
 \chi_{3}
   &=\sum\limits_{p}\frac{J_{p}(\frac{F}{\omega})J_{-p}(\frac{F}{\omega})}{p-m+m'+u'}, \nonumber
    \\  \chi_{4}
   &=\sum\limits_{p}\frac{J_{p}^{2}(\frac{F}{\omega})}{-p-m+m'+u'},  \nonumber
   \\  \chi_{5}
   &=\sum\limits_{p}\frac{J_{p}(\frac{F}{\omega})J_{-p}(\frac{F}{\omega})}{-p+m+m'+u'}, \nonumber
    \\  \chi_{6}
   &=\sum\limits_{p}\frac{J_{p}^{2}(\frac{F}{\omega})}{p+m+m'+u'}.
\end{align}
From equation (\ref{order2}), we observe that the dynamics of a particle initially occupying the impurity site is frozen in the second-order approximation,
whereas the transition $|-1,\uparrow\rangle\rightarrow |1,\uparrow\rangle$ (or $|-1,\downarrow\rangle\rightarrow |1,\downarrow\rangle$) is generated by a second-order transition process via the virtual intermediate state $|0,\downarrow\rangle$ (or  $|0,\uparrow\rangle$).
This second-order transition describes such a physical process, in which a single spin-up (down) particle can only tunnel back and forth between the two nearest-neighbor sites of the impurity in a manner that the spin is nonetheless kept unchanged. This spin-invariant behavior exhibits remarkable resemblance to the usual spin-conserving tunneling case.
However, in contrast to the usual spin-conserving tunneling case, the second-order tunneling process described by equation (\ref{order2}) is essentially spin-dependent and has distinct physical mechanisms, which can be viewed from the fact that  two types of second-order transition processes $|-1,\uparrow\rangle\rightarrow|0,\downarrow\rangle\rightarrow |1,\uparrow\rangle$ and  $|-1,\downarrow\rangle\rightarrow|0,\uparrow\rangle\rightarrow |1,\downarrow\rangle$ are generated through elimination of different virtual intermediate states ($|0,\downarrow\rangle$ or $|0,\uparrow\rangle$) and thus have distinct effective tunneling rates  given by $\omega\epsilon^2\chi_{3}$ and $\omega\epsilon^2\chi_{5}$ respectively.
Clearly, the tunneling dynamics (circles in Fig.~{\ref{figure5} (b)) obtained from the second-order perturbative results (equation ({\ref{order2})) are
in consistency with those on the basis of a full numerical
analysis of the original model (\ref {Hamiltonian}) for the purely spin-flipping tunneling ($\cos\alpha=0$) case, see the curves in Fig.~{\ref{figure5} (b).

Following the same procedure as in subsection A, we can analytically calculate the quasienergies corresponding to the purely spin-flipping tunneling ($\cos\alpha=0$) case under DL condition in the far-off-resonant regime. Substituting $A_{n,\sigma}(t)=A'_{n,\sigma}\exp(-iEt)$ into Eq.~(\ref{order2}),
we have the resulting time-independent form of equation (\ref{order2}) which gives the eigenvector components  $A'_{n,\sigma}$ and the eigenvalue $E$.
With the help of Eqs.~(\ref{subst}) and (\ref{subst2}),
  the analytical Floquet solutions of the considered system  can be constructed as $a_{n,\sigma}=\tilde{a}_{n,\sigma}(t)\exp(-i\varepsilon t)=A_{n,\sigma}(t)\exp\{-i[n\Phi(t)+(-1)^k\frac{\Omega}{2}t+\varepsilon_{0}\delta_{n,0}t]\}=A'_{n,\sigma}\exp\{-i[n\Phi(t)+(-1)^k\frac{\Omega}{2}t+\varepsilon_{0}\delta_{n,0}t]\}\exp(-iEt)$ ($k=0,1$ for $\sigma=\uparrow,\downarrow$, respectively). Imposing the $T$-periodic boundary condition on these
states $\tilde{a}_{n,\sigma}(t)$
reveals the corresponding
quasienergies (modulo $\omega$) to be
\begin{align}
\label{Qeor3}
\varepsilon_{1}&=\frac{-v^2\chi_4+v^2\chi_3}{\omega}+\frac{\omega}{2},\nonumber\\
    \varepsilon_{2}&=\frac{2v^2\chi_4}{\omega}+u+\frac{\omega}{2},\nonumber\\
    \varepsilon_{3}&=\frac{-v^2\chi_4-v^2\chi_3}{\omega}+\frac{\omega}{2},\nonumber\\
    \varepsilon_{4}&=\frac{-v^2\chi_6+v^2\chi_5}{\omega}+\frac{\omega}{2},\nonumber\\
    \varepsilon_{5}&=\frac{2v^2\chi_6}{\omega}+u+\frac{\omega}{2},\nonumber\\
    \varepsilon_{6}&=\frac{-v^2\chi_6-v^2\chi_5}{\omega}+\frac{\omega}{2},\nonumber\\
\end{align}
where the identities $\Omega=m\omega,\varepsilon_{0}=m'\omega+u$, $m, m'=1,2,...$, have been used.

\begin{figure}[htb]
\includegraphics[width=0.5\textwidth]{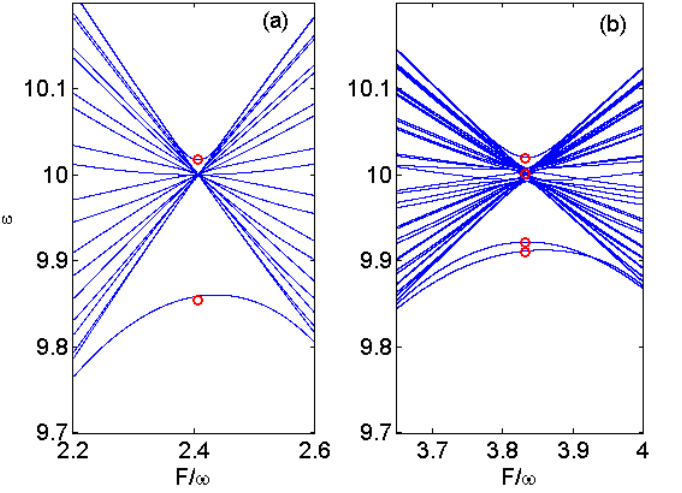}
\caption{(Color online) Comparison between the numerically computed quasienegies (solid lines) from the original model (\ref{Hamiltonian}) and the analytical quasienegies (open circles) based on  the second-order perturbative results (equations (\ref{Qeor2}) and (\ref{Qeor3})) for the nonresonant case $\varepsilon_0/\omega=1.2$, at (a) $\sin\alpha=0$ and (b) $\cos\alpha=0$.  The other parameters are $v=1, \omega=20, \Omega=20$. For clarity, we have only
plotted a portion of the quasienergy levels.}\label{figure6}
\end{figure}

Our analytical quasienergies based on the second-order perturbative results of equations ({\ref{Qeor2}) and ({\ref{Qeor3})
are plotted as open circles in Figs.~\ref{figure6} (a) and (b), corresponding to the usual spin-conserving tunneling ($\sin\alpha=0$) and
the purely spin-flipping tunneling ($\cos\alpha=0$) case respectively. For clarity, we have only
plotted a portion of the quasienergy levels. The two quasienergies (the analytical correspondences $\varepsilon_{2,\sigma}$ in equation ({\ref{Qeor2}) and $\varepsilon_2, \varepsilon_5$ in  equation ({\ref{Qeor3}) ),
not depicted here, are separated widely from the rest of the band,
corresponding to the localized Floquet states in which the particle is captured by the impurity. It is clearly shown that the other four analytical quasienergies given by equations ({\ref{Qeor2}) and ({\ref{Qeor3}) match very well
with the direct numerical results obtained from the original model (\ref {Hamiltonian}). As shown in Fig.~\ref{figure6}, the four analytical quasienergies for
the usual spin-conserving tunneling case are doubly degenerate and this degeneracy
breaks up for the purely spin-flipping tunneling case.
Under the DL conditions, all the other quasienergies collapse into one single value (the value is $\omega/2$ for our selected parameters), and thus the analytical quasienergies (see open circles in Fig.~\ref{figure6}) again verify the existence of avoided level crossings for the off-resonant case with moderate reduced
impurity strength $u$.

\section{Discussion and experimental aspects}
Here, we first justify the validity of the effective three-site model approximation and the multiple-time-scale asymptotic perturbative approximation used in this work.
To this end, we define two related variables: one is the time-averaged total probability of finding the single particle in sites $0, \pm 1$, $\langle S_1\rangle=\frac{1}{\Delta t}\int_{0}^{\Delta t}\sum\limits_{\sigma}(P_{-1,\sigma}+P_{0,\sigma}+P_{1,\sigma})dt$; the other is the time-averaged total probability of finding the single particle in sites $\pm 1$, $\langle S_2\rangle=\frac{1}{\Delta t}\int_{0}^{\Delta t}\sum\limits_{\sigma}(P_{-1,\sigma}+P_{1,\sigma})dt$. The former is used to quantify the validity of the effective three-site model approximation: if
$\langle S_1\rangle\approx 1$ lasts  for a long-enough averaging time interval $\Delta t$, the effective three-site model approximation is valid, otherwise it breaks down.
The latter is to measure the validity of the multiple-time-scale asymptotic analysis. If the average of the occupation probabilities of sites $\pm 1$ over a long-enough time interval $\Delta t$ is approximately equal to one, $\langle S_2\rangle\approx 1$, it means that the single atom only
tunnels between the two nearest-neighbor
sites of the impurity with negligible impurity site population. When $\langle S_2\rangle\approx 1$, the second-order tunneling occurs and
the validity of the multiple-time-scale asymptotic analysis can be justified. To delineate the validity regime
of the effective three-site model approximation and the multiple-time-scale asymptotic perturbative approximation, we numerically give $\langle S_1\rangle$ (solid lines) and $\langle S_2\rangle$ (dashed lines)  as the function of the driving frequency $\omega$, for the usual spin-conserving tunneling ($\sin\alpha=0$) case with the parameters $v=1,\varepsilon_0=1.2\omega, \Omega=\omega, F/\omega=2.405$, as in figure \ref{figure7} (a), and for the purely spin-flipping tunneling ($\cos\alpha=0$) case with the parameters $v=1,\varepsilon_0=1.2\omega, \Omega=\omega, F/\omega=3.8317$, as in figure \ref{figure7} (b).
In our numerical calculations we have always assumed that the system is initialized in $|-1,\uparrow\rangle$ and we have shown the second
variable $\langle S_2\rangle$ by actually computing $\langle S_2\rangle=\frac{1}{\Delta t}\int_{0}^{\Delta t}(P_{-1,\uparrow}+P_{1,\uparrow})dt$ for $\Delta t=200(v^{-1})$ because of $P_{-1,\downarrow}=P_{1,\downarrow}=0$.
It can be seen from Fig.~\ref{figure7} that the driving frequency $\omega$ on the order of or larger than $10v$ can be regarded  safe for observing our theoretical predictions.

\begin{figure}[htb]
\includegraphics[width=0.5\textwidth]{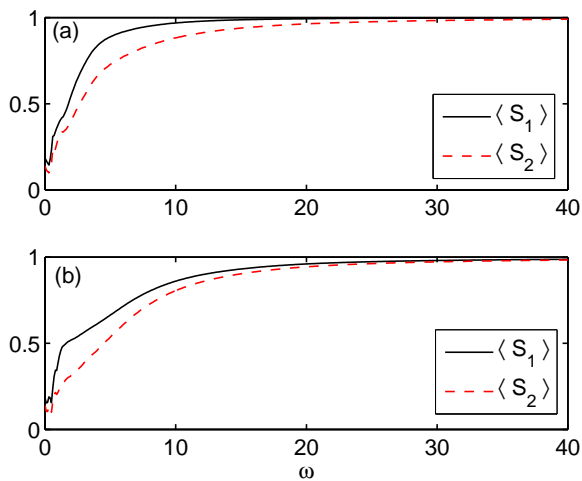}
\caption{(Color online) Two quantities $\langle S_1\rangle$ and $\langle S_2\rangle$ versus the driving frequency $\omega$ with the system initialized in state $|-1,\uparrow\rangle$. Here, $\langle S_1\rangle$ denotes the time-averaged total probability of finding the single particle in sites $0, \pm 1$, and $\langle S_2\rangle$ denotes the time-averaged total probability of finding the single spin-up particle in sites $\pm 1$. (a) $\sin\alpha=0$, $F/\omega=2.405$; and (b) $\cos\alpha=0$, $F/\omega=3.8317$. The time used for averaging is $200$ in dimensionless unit. In the two plots, we have set $v=1, \varepsilon_0=1.2\omega, \Omega=\omega$ and
the size of the lattice to be $N = 21$.}\label{figure7}
\end{figure}

In the numerical calculations above, we have set that the size of the
lattice is $N=21$. Our numerical results (not shown) demonstrate that
 the magnitude of the driving frequency for validity of both approximations employed will
decrease with the decrease of lattice size, and the effective three-site model approximation and the second-order perturbative approximation  are still applicable even when the
 driving frequency is much smaller than $10v$ if we treat relatively small systems of up to $N>3$.
 This indicates that our results are
more realistic in the systems of up to relatively small numbers of sites, because the validity of  single-band tight-binding description of
the original model  requires moderately small driving frequency.

Before concluding, we present some remarks on experimental aspects of our
theoretical predictions. In our work, we only address the tunneling dynamics of the usual spin-conserving tunneling ($\sin\alpha=0$) and
the purely spin-flipping tunneling ($\cos\alpha=0$) case. For the coexistence of the spin-conserving tunneling
and the spin-flipping tunneling case, the simultaneous dynamical localization does not occur, because the collapse of these two quasienergy
bands corresponding to the couplings between states with the same (different) spins respectively occurs at distinct values of the scaled driving amplitude $F/\omega$. It renders this problem
 more complicated, which deserves further studies in future.
As already noted, possible experimental identification of our
theoretical predictions requires precise control of the dimensionless SO-coupling strength $\alpha=\pi k_{r}/k_{\rm{lat}}$.
In experiment, the SOC strength is determined by the wavelength
of the Raman beams and the angle at which they intersect. More recently, the techniques for tuning SO coupling
have been successfully realized in experiments\cite{Spielman,You}: for example, the SO-coupling strength can be tuned by the method of shaking\cite{YZhang,Spielman}. In addition, the momentum-independent
Zeeman field can be produced by the Raman beams and the parameter $\Omega$ can be tuned by changing the intensity of Raman beams.
With the growing state-of-the-art experimental creation of artificial SO-coupled bosonic gases in an optical lattice, we hope the present work can contribute to the detailed checks of our current understanding of the novel tunneling dynamics of a SO-coupled particle loaded into a periodically driven optical lattice with an impurity.

\section{Conclusion}
In summary, we have identified two types of second-order tunneling process for a single SO-coupled atom placed in a driven optical lattice with an impurity, provided that the impurity potential strength is in far-off-resonance
with the driving field. The two types of second-order tunneling seem to show up in the same way that an initially spin-up (down) atom
only tunnels between the two nearest-neighbor
sites of the impurity, in which the spin remains unchanged and the impurity site population is negligible during all the evolution time.
Nevertheless, it has been found that two distinct mechanisms are responsible for the two second-order tunneling processes: one is related to a spin-independent tunneling process in which the virtual intermediate state without spin-flipping is eliminated, the other is to a spin-dependent tunneling process in which the second-order transition is generated via elimination of a virtual intermediate state with spin-flipping.
Fortunately, the two types of second-order tunneling processes can be distinguished by tuning the ratio of the impurity potential to the driving frequency.
When the ratio becomes an integer, the spin-independent second-order tunneling is tuned to the resonant oscillation without spin-flipping between the impurity and its two nearest-neighbor sites, while  the spin-dependent second-order tunneling is to the resonant oscillation with spin-flipping.

By means of the direct quasienergy-band computation, we evidence that all the resonant and  second-order tunneling processes are reflected by the avoided level crossings
near the pseudocollapse points of quasienergy bands. Furthermore, the very subtle and fine avoided level crossing in the
quasienergy spectrum, which manifests itself dynamically in the second-order tunneling and is easily
neglected in exact numerical simulations,  has been fully understood in the
framework of the second-order perturbative theory. These results may be relevant to engineering the spin-dependent quantum transport in experiments and
find some possible applications in the design of novel spintronics devices.

\begin{acknowledgments}
The work is funded by the NNSF of China under Grants 11465009, 11747172,
11165009, and Humanity and Social Science Youth foundation of Ministry of Education of China (17YJCZH042). Xiaobing Luo thanks Zheng Zhou for helpful discussions,
and also feels deeply indebted to Shoulin Feng for her supports and encouragements during the completion of this work. Yu Guo is supported by Hunan Provincial Natural
Science Foundation of China under grant No. 2017JJ2272, the Open Research Fund of the Hunan Province Higher Education Key Laboratory of Modeling and Monitoring on the Near-Earth Electromagnetic Environments, Changsha University of Science and Technology under No. 20150108, and the Opening Project of Key Laboratory of Low Dimensional Quantum Structures and Quantum Control of Ministry of Education under Grant No. QSQC1402.
\end{acknowledgments}
\appendix

\section{}
\setcounter{equation}{0}
\renewcommand{\theequation}{A.\arabic{equation}}
In Appendix A, we will give the detailed derivations to equation (\ref{order1}) in the main text for the amplitudes $A_{n,\sigma}$ ($n=0,\pm 1$) up to the second-order $\epsilon^{2}$ for the usual spin-conserving tunneling ($\sin\alpha=0$) case. In the high-frequency regime and when the DL condition $J_0(F/\omega)=0$ is satisfied, truncating the Hilbert
space to two independent subspaces: $\{|-1 ,\uparrow\rangle,|0 ,\uparrow\rangle,|1 ,\uparrow\rangle\}$ and $\{|-1 ,\downarrow\rangle,|0 ,\downarrow\rangle,|1 ,\downarrow\rangle\}$ is assumed to produce an effective three-site model. Under such assumptions, taking $\epsilon\equiv v/\omega$ as a small
parameter and introducing the normalized time
variable $\tau = \omega t$, we rewrite equation (\ref{NAmPro}) in the truncated space as
\begin{align}\label{NAmPro2}
     i\frac{d b_{-1,\sigma}}{d\tau}=&-\epsilon\cos\alpha [{b}_{0,\uparrow}e^{-i\Phi(\tau)-i(m'+u') \tau}],\nonumber\\
     i\frac{d b_{0,\sigma}}{d\tau}=&-\epsilon\cos\alpha [{b}_{1,\sigma}e^{-i\Phi(\tau)+i(m'+u') \tau}\nonumber\\ &+{b}_{-1,\sigma}e^{i\Phi(\tau)+i(m'+u')\tau}],\nonumber\\
     i\frac{d b_{1,\sigma}}{d\tau}=&-\epsilon\cos\alpha [{b}_{0,\sigma}e^{i\Phi(\tau)-i(m'+u') \tau}],
    \end{align}
where we have set $u'=u/\omega$.
Let us look for the solution $b_{n,\sigma} (n=-1,0,1, \sigma=\uparrow, \downarrow)$ to Eq.~(\ref{NAmPro2}) as a power-series expansion
in the smallness parameter $\epsilon$
\begin{equation}\label{powerseriesA}
    b_{n,\sigma}(\tau)=b_{n,\sigma}^{(0)}(\tau)+\epsilon b_{n,\sigma}^{(1)}(\tau)+\epsilon^{2} b_{n,\sigma}^{(2)}(\tau)+\cdot\cdot\cdot,
\end{equation}
and Let us introduce multiple time scales $\tau_{k}=\epsilon^k\tau, k=0,1,2,...$.

By using the derivative rule $d/d\tau=\partial_{\tau_{0}}+\epsilon\partial_{\tau_{1}}+\epsilon^{2}\partial_{\tau_{2}}+\cdot\cdot\cdot$, and substituting
equation (\ref{powerseriesA}) into equation (\ref{NAmPro2}), we obtain
 a hierarchy of equations for successive corrections to
$b_{n,\sigma} (n=-1,0,1, \sigma=\uparrow, \downarrow)$ at different orders in $\epsilon$. In the calculation process, we take $\varepsilon_0$ far from any integer multiple of $\omega$, namely, the reduced impurity potential $|u|$ is sufficiently large.
At the leading order $\epsilon^{0}$, we find
\begin{equation}\label{zero1}
\partial_{\tau_{0}}b_{n,\sigma}^{(0)}(\tau)=0,~~~ b_{n,\sigma}^{(0)}(\tau)=A_{n,\sigma}(\tau_{1},\tau_{2},\cdot\cdot\cdot),
\end{equation}
where the amplitudes $A_{n,\sigma}(\tau_{1},\tau_{2},\cdot\cdot\cdot)$ are functions of the slow time variables $\tau_{1},\tau_{2},...$, but
independent of the fast time variable $\tau_{0}$.
At order $\epsilon^{1}$, one obtains
\begin{align}\label{MTorder11}
     i\frac{\partial
     b_{-1,\sigma}^{(1)}}{\partial\tau_0}=&-i\partial_{\tau_{1}}A_{-1,\sigma}-\cos\alpha [{A}_{0,\sigma}e^{-i\Phi(\tau_0)-i(m'+u') \tau_0}],\nonumber\\
     i\frac{\partial b_{0,\sigma}^{(1)}}{\partial\tau_0}=&-i\partial_{\tau_{1}}A_{0,\sigma}-\cos\alpha [{A}_{1,\sigma}e^{-i\Phi(\tau_0)+i(m'+u') \tau_0}\nonumber\\&+{A}_{-1,\sigma}e^{i\Phi(\tau_0)+i(m'+u')\tau_0}],\nonumber\\
    i\frac{\partial b_{1,\sigma}^{(1)}}{\partial\tau_0}=&-i\partial_{\tau_{1}}A_{1,\sigma}-\cos\alpha [{A}_{0,\sigma}e^{i\Phi(\tau_0)-i(m'+u') \tau_0}].
\end{align}
For the convenience of our discussion, we simplify equation (\ref{MTorder11}) as
\begin{equation}
  i\frac{\partial b_{n,\sigma}^{(1)}}{\partial\tau_0}=-i\partial_{\tau_{1}}A_{n,\sigma}+K^{(1)}_{n,\sigma}(\tau_0).
\end{equation}
 To avoid the
occurrence of secularly growing terms in the solution $b_{n,\sigma}^{(1)}$, the solvability condition
\begin{equation}\label{order1(2)}
    i\partial_{\tau_{1}}A_{n,\sigma}=\overline{K^{(1)}_{n,\sigma}(\tau_0)}
\end{equation}
must be satisfied. Throughout our paper, the overline denotes the time average with respect to the fast time
variable $\tau_0$.
Obviously, we have
\begin{equation}\label{order1(3)}
i\partial_{\tau_{1}}A_{n,\sigma}=\overline{K^{(1)}_{n,\sigma}(\tau_0)}=0,
\end{equation}
which represents that the dc component of the driving term $K^{(1)}_{n,\sigma}(\tau_0)$ is zero. According to $b_{n,\sigma}^{(1)}=-i\int [K^{(1)}_{n,\sigma}-\overline{K^{(1)}_{n,\sigma}(\tau_0)}]d\tau_0$, the amplitudes $b_{n,\sigma}$ at order $\epsilon$ are given by
\begin{align}\label{solutionorder1(2)}
   b_{-1,\sigma}^{(1)}&=-\cos\alpha  F_{0}^{\ast}(\tau_{0}){A}_{0,\sigma},\nonumber
      \\ b_{0,\sigma}^{(1)}&=\cos\alpha[F_{1}(\tau_{0}){A}_{1,\sigma}+F_{0}(\tau_{0}){A}_{-1,\sigma}],\nonumber
   \\ b_{1,\sigma}^{(1)}&=-\cos\alpha F_{1}^{\ast}(\tau_{0}){A}_{0,\sigma},
\end{align}
where
\begin{align}
    F_{0}(\tau_{0})
    &=\sum\limits_{p}\frac{J_{p}(\frac{F}{\omega})e^{i (p+m'+u')\tau_{0}}}{p+m'+u'},\nonumber\\
   F_{1}(\tau_{0})
    &=\sum\limits_{p}\frac{J_{p}(\frac{F}{\omega})e^{-i (p-m'-u')\tau_{0}}}{-(p-m'-u')}.
    \end{align}
At the next order $\epsilon^{2}$, we find
\begin{equation}
  i\frac{\partial b_{n,\sigma}^{(2)}}{\partial\tau_0}=-i\partial_{\tau_{2}}A_{n,\sigma}-i\partial_{\tau_{1}} b_{n,\sigma}^{(1)}+K^{(2)}_{n,\sigma}(\tau_0),
\end{equation}
with
\begin{align}\label{MTorder12}
    K^{(2)}_{-1,\sigma}(\tau_0)=&-\cos\alpha [{b}_{0,\sigma}^{(1)}e^{-i\Phi(\tau_0)-i(m'+u') \tau_0}],\nonumber\\
     K^{(2)}_{0,\sigma}(\tau_0)=&-\cos\alpha [{b}_{1,\sigma}^{(1)}e^{-i\Phi(\tau_0)+i(m'+u') \tau_0}\nonumber\\&+{b}_{-1,\sigma}^{(1)}e^{i\Phi(\tau_0)+i(m'+u')\tau_0}],\nonumber\\
    K^{(2)}_{1,\sigma}(\tau_0)=&-\cos\alpha [{b}_{0,\sigma}^{(1)}e^{i\Phi(\tau_0)-i(m'+u') \tau_0}].
\end{align}
In order to avoid the occurrence of secularly growing terms in solution $b_{n,\sigma}
^{(2)}$, the following
solvability condition must be satisfied:
\begin{align}\label{order2{2}}
    i\partial_{\tau_{2}}A_{-1,\sigma}&=\overline{-i\frac{\partial b_{-1,\sigma}^{(1)}}{\partial\tau_1}+K^{(2)}_{-1,\sigma}}=- (\chi_{1}A_{1,\sigma}+\chi_{2}A_{-1,\sigma})\nonumber
    \\i\partial_{\tau_{2}}A_{0,\sigma}&=\overline{-i\frac{\partial b_{0,\sigma}^{(1)}}{\partial\tau_1}+K^{(2)}_{0,\sigma}}=2\chi_{2}A_{0,\sigma}\nonumber
    \\i\partial_{\tau_{2}}A_{1,\sigma}&=\overline{-i\frac{\partial b_{1,\sigma}^{(1)}}{\partial\tau_1}+K^{(2)}_{1,\sigma}}=-(\chi_{2}A_{1,\sigma}+\chi_{1}A_{-1,\sigma}),
\end{align}
where we have set
\begin{align}
  \chi_{1}
   &=\sum\limits_{p}\frac{J_{p}(\frac{F}{\omega})J_{-p}(\frac{F}{\omega})}{-p+m'+u'}, \nonumber
    \\  \chi_{2}&=\sum\limits_{p}\frac{J_{p}^{2}(\frac{F}{\omega})}{p+m'+u'}.
\end{align}
Thus the evolution of the amplitudes $A_{n,\sigma}$ ($n=0,\pm 1$) up to the second-order long time scale is given by
\begin{equation}\label{allorder}
  \frac{dA_{n,\sigma}}{d\tau}=(\frac{\partial}{\partial\tau_0}+\epsilon\frac{\partial}{\partial\tau_1}+\epsilon^2\frac{\partial}{\partial\tau_2})A_{n,\sigma}.
\end{equation}
Substituting equation (\ref{zero1}), equation (\ref{order1(3)}) and (\ref{order2{2}}) into equation (\ref{allorder}), we obtain
the set of coupled equations, equation (\ref{order1}) in the main text.

\section{}
\setcounter{equation}{0}
\renewcommand{\theequation}{B.\arabic{equation}}
This Appendix will give the derivation of the evolution equation (\ref{order2}) for the amplitudes $A_{n,\sigma}$ ($n=0,\pm 1$)
on the  second-order long  time scale for the purely spin-flipping tunneling ($\cos\alpha=0$) case.  In
the high-frequency limit and under the DL condition, the system described by
 Hamiltonian (\ref{Hamiltonian}) can be truncated into the effective
three-site system with the dynamics confined in two independent subspaces: $\{|-1,\uparrow\rangle, |0,\downarrow\rangle, |1,\uparrow\rangle\}$ and $\{|-1,\downarrow\rangle, |0,\uparrow\rangle, |1,\downarrow\rangle\}$. In the frame of effective three-site model, let $\tau=\omega t, \epsilon=\upsilon/\omega, u'=u/\omega$, equation (\ref{NAmPro}) can be rewritten as
\begin{align}\label{NAmPro1}
     i\frac{d b_{-1,\uparrow}}{d\tau}=&-\epsilon\sin\alpha [-{b}_{0,\downarrow}e^{-i\Phi(\tau)+i(m-m'-u') \tau}],\nonumber\\
     i\frac{d b_{0,\downarrow}}{d\tau}=&-\epsilon\sin\alpha [{b}_{1,\uparrow}e^{-i\Phi(\tau)-i(m-m'-u') \tau}\nonumber\\&-{b}_{-1,\uparrow}e^{i\Phi(\tau)-i(m-m'-u') \tau}],\nonumber\\
     i\frac{d b_{1,\uparrow}}{d\tau}=&-\epsilon\sin\alpha [{b}_{0,\downarrow}e^{i\Phi(\tau)+i(m-m'-u') \tau}],\nonumber\\
    i\frac{d b_{-1,\downarrow}}{d\tau}=&-\epsilon\sin\alpha [{b}_{0,\uparrow}e^{-i\Phi(\tau)-i(m+m'+u') \tau}],\nonumber\\
     i\frac{d b_{0,\uparrow}}{d\tau}=&-\epsilon\sin\alpha [-{b}_{1,\downarrow}e^{-i\Phi(\tau)+i(m+m'+u') \tau}\nonumber\\&+{b}_{-1,\downarrow}e^{i\Phi(\tau)+i(m+m'+u') \tau}],\nonumber\\
     i\frac{d b_{1,\downarrow}}{d\tau}=&-\epsilon\sin\alpha [-{b}_{0,\uparrow}e^{i\Phi(\tau)-i(m+m'+u') \tau}].
\end{align}

Following the same procedure as outlined in Appendix A,  we perform a multiple-scale asymptotic
analysis of equation (\ref{NAmPro1}) by introducing the multiple-time-scale variables $\tau_{k}=\epsilon^k\tau, k=0,1,2,...$ and writing the solution as an expansion in powers of $\epsilon$, and thus derive a hierarchy of approximation equations of different orders in $\epsilon$.
At the leading order $\epsilon^0$, we also have
\begin{equation}\label{zero1b}
\partial_{\tau_{0}}b_{n,\sigma}^{(0)}(\tau)=0,~~~ b_{n,\sigma}^{(0)}(\tau)=A_{n,\sigma}(\tau_{1},\tau_{2},\cdot\cdot\cdot).
\end{equation}
At the next order $\epsilon^{1}$, we arrive at the coupled equations
\begin{equation}\label{lb}
  i\frac{\partial b_{n,\sigma}^{(1)}}{\partial\tau_0}=-i\partial_{\tau_{1}}A_{n,\sigma}+G^{(1)}_{n,\sigma}(\tau_0),
\end{equation}
with
\begin{align}\label{MTorder1b}
     G^{(1)} _{-1,\uparrow}=&-\sin\alpha [-{A}_{0,\downarrow}e^{-i\Phi(\tau_0)+i(m-m'-u') \tau_0}],\nonumber\\
     G^{(1)}_{0,\downarrow}=&-\sin\alpha [{A}_{1,\uparrow}e^{-i\Phi(\tau_0)-i(m-m'-u') \tau_0}\nonumber\\&-{A}_{-1,\uparrow}e^{i\Phi(\tau_0)-i(m-m'-u') \tau_0}],\nonumber\\
        G^{(1)}_{1,\uparrow}=&-\sin\alpha [{A}_{0,\downarrow}e^{i\Phi(\tau_0)+i(m-m'-u') \tau_0}],\nonumber\\
     G^{(1)}_{-1,\downarrow}=&-\sin\alpha [{A}_{0,\uparrow}e^{-i\Phi(\tau_0)-i(m+m'+u') \tau_0}],\nonumber\\
      G^{(1)}_{0,\uparrow}=&-\sin\alpha [-{A}_{1,\downarrow}e^{-i\Phi(\tau_0)+i(m+m'+u') \tau_0}\nonumber\\&+{A}_{-1,\downarrow}e^{i\Phi(\tau_0)+i(m+m'+u') \tau_0}],\nonumber\\
     G^{(1)}_{1,\downarrow}=&-\sin\alpha [-{A}_{0,\uparrow}e^{i\Phi(\tau_0)-i(m+m'+u') \tau_0}].
\end{align}
 The avoidance of secularly growing terms in $b_{n,\sigma}^{(1)}$ requires
 \begin{equation}\label{order11b}
    i\partial_{\tau_{1}}A_{n,\sigma}=\overline{ G^{(1)}_{n,\sigma}},
\end{equation}
where the average value $\overline{G^{(1)}_{n,\sigma}}$ over the fast time variable $\tau_0$ denotes the dc component of the driving term $G^{(1)}_{n,\sigma}$.
According to the expressions (\ref{MTorder1b}), it is obvious that
\begin{equation}\label{avoid}
    i\partial_{\tau_{1}}A_{n,\sigma}=0.
\end{equation}
The correction of $b_{n,\sigma}$  at order
$\epsilon^{1}$ can be then calculated as
\begin{equation}\label{amplitudelb}
   b_{n,\sigma}^{(1)}=-i\int( G^{(1)}_{n,\sigma}- \overline{G^{(1)}_{n,\sigma}})d\tau,
\end{equation}
and the solutions of order
$\epsilon^{1}$ read
\begin{align}\label{solutionorder1b}
   b_{-1,\uparrow}^{(1)}&=\sin\alpha F_{2}^{\ast}(\tau_{0}){A}_{0,\downarrow},\nonumber
   \\ b_{0,\downarrow}^{(1)}&=\sin\alpha [F_{3}(\tau_{0}){A}_{1,\uparrow}-F_{2}(\tau_{0}){A}_{-1,\uparrow}],\nonumber
   \\ b_{1,\uparrow}^{(1)}&=-\sin\alpha F_{3}^{\ast}(\tau_{0}){A}_{0,\downarrow},\nonumber
   \\ b_{-1,\downarrow}^{(1)}&=-\sin\alpha F_{4}^{\ast}(\tau_{0}){A}_{0,\uparrow}\nonumber
      \\ b_{0,\uparrow}^{(1)}&=\sin\alpha [-F_{5}(\tau_{0}){A}_{1,\downarrow}+F_{4}(\tau_{0}){A}_{-1,\downarrow}],\nonumber
    \\b_{1,\downarrow}^{(1)}&=\sin\alpha F_{5}^{\ast}(\tau_{0}){A}_{0,\uparrow},
\end{align}
where
\begin{align}
   F_{2}(\tau_{0})
    &=\sum\limits_{p}\frac{J_{p}(\frac{F}{\omega})e^{i (p-m+m'+u')\tau_{0}}}{p-m+m'+u'},\nonumber \\
   F_{3}(\tau_{0})
    &=\sum\limits_{p}\frac{J_{p}(\frac{F}{\omega})e^{-i (p+m-m'-u')\tau_{0}}}{-(p+m-m'-u')},\nonumber \\
    F_{4}(\tau_{0})
    &=\sum\limits_{p}\frac{J_{p}(\frac{F}{\omega})e^{i (p+m+m'+u')\tau_{0}}}{p+m+m'+u'},\nonumber \\
    F_{5}(\tau_{0})
    &=\sum\limits_{p}\frac{J_{p}(\frac{F}{\omega})e^{-i (p-m-m'-u')\tau_{0}}}{-(p-m-m'-u')}.
\end{align}
Next, we consider the asymptotic
analysis up to the order $\epsilon^{2}$:
\begin{equation}\label{2b}
  i\frac{\partial b_{n,\sigma}^{(2)}}{\partial\tau_0}=-i\partial_{\tau_{2}}A_{n,\sigma}-i\partial_{\tau_{1}} b_{n,\sigma}^{(1)}+G^{(2)}_{n,\sigma},
\end{equation}
where
\begin{align}\label{MTorder2b}
   G^{(2)}_{-1,\uparrow}=&-\sin\alpha [-{b}_{0,\downarrow}^{(1)}e^{-i\Phi(\tau_0)+i(m-m'-u') \tau_0}],\nonumber\\
      G^{(2)}_{0,\downarrow}=&-\sin\alpha [{b}_{1,\uparrow}^{(1)}e^{-i\Phi(\tau_0)-i(m-m'-u') \tau_0}\nonumber\\&-{b}_{-1,\uparrow}^{(1)}e^{i\Phi(\tau_0)-i(m-m'-u') \tau_0}],\nonumber\\
      G^{(2)}_{1,\uparrow}=&-\sin\alpha [{b}_{0,\downarrow}^{(1)}e^{i\Phi(\tau_0)+i(m-m'-u') \tau_0}],\nonumber\\
      G^{(2)}_{-1,\downarrow}=&-\sin\alpha [{b}_{0,\uparrow}^{(1)}e^{-i\Phi(\tau_0)-i(m+m'+u') \tau_0}],\nonumber\\
       G^{(2)}_{0,\uparrow}=&-\sin\alpha [-{b}_{1,\downarrow}^{(1)}e^{-i\Phi(\tau_0)+i(m+m'+u') \tau_0}\nonumber\\&+{b}_{-1,\downarrow}^{(1)}e^{i\Phi(\tau_0)+i(m+m'+u') \tau_0}],\nonumber\\
       G^{(2)}_{1,\downarrow}=&-\sin\alpha [-{b}_{0,\uparrow}^{(1)}e^{i\Phi(\tau_0)-i(m+m'+u') \tau_0}].
\end{align}
Given the fact that $\overline{-i\partial_{\tau_{1}} b_{n,\sigma}^{(1)}}=0$, the solvability condition at order $\epsilon^2$ then yields
\begin{align}\label{order2b}
    i\partial_{\tau_{2}}A_{-1,\uparrow}&=\overline{G^{(2)}_{-1,\uparrow}}=\chi_{3}A_{1,\uparrow}-\chi_{4}A_{-1,\uparrow},\nonumber
    \\i\partial_{\tau_{2}}A_{0,\downarrow}&=\overline{G^{(2)}_{0,\downarrow}}=2\chi_{4}A_{0,\downarrow},\nonumber
    \\i\partial_{\tau_{2}}A_{1,\uparrow}&=\overline{G^{(2)}_{1,\uparrow}}=-(\chi_{4}A_{1,\uparrow}-\chi_{3}A_{-1,\uparrow}),\nonumber
     \\i\partial_{\tau_{2}}A_{-1,\downarrow}&=\overline{G^{(2)}_{-1,\downarrow}}=\chi_{5}A_{1,\downarrow}-\chi_{6}A_{-1,\downarrow},\nonumber
    \\i\partial_{\tau_{2}}A_{0,\uparrow}&=\overline{G^{(2)}_{0,\uparrow}}=2\chi_{6}A_{0,\uparrow},\nonumber
    \\i\partial_{\tau_{2}}A_{1,\downarrow}&=\overline{G^{(2)}_{1,\downarrow}}=-(\chi_{6}A_{1,\downarrow}-\chi_{5}A_{-1,\downarrow}),
\end{align}
where we have set
\begin{align}
  \chi_{3}
   &=\sum\limits_{p}\frac{J_{p}(\frac{F}{\omega})J_{-p}(\frac{F}{\omega})}{p-m+m'+u'}, \nonumber
    \\  \chi_{4}
   &=\sum\limits_{p}\frac{J_{p}^{2}(\frac{F}{\omega})}{-p-m+m'+u'},  \nonumber
   \\  \chi_{5}
   &=\sum\limits_{p}\frac{J_{p}(\frac{F}{\omega})J_{-p}(\frac{F}{\omega})}{-p+m+m'+u'}, \nonumber
    \\  \chi_{6}
   &=\sum\limits_{p}\frac{J_{p}^{2}(\frac{F}{\omega})}{p+m+m'+u'}.
\end{align}
To sum up, the evolution of the amplitudes $A_{n,\sigma}$ ($n=0,\pm 1$) up to the second-order long time scale is given by
\begin{equation}\label{allorderb}
  \frac{dA_{n,\sigma}}{d\tau}=(\frac{\partial}{\partial\tau_0}+\epsilon\frac{\partial}{\partial\tau_1}+\epsilon^2\frac{\partial}{\partial\tau_2})A_{n,\sigma}.
\end{equation}
Substitution of equations (\ref{zero1b}), (\ref{avoid}) and (\ref{order2b}) into equation (\ref{allorderb}) yields the evolution equation (\ref{order2}) for the amplitudes $A_{n,\sigma}$ ($n=0,\pm 1$)
on the  second-order long  time scale for the purely spin-flipping tunneling ($\cos\alpha=0$) case.

\bibliography{basename of .bib file}

\begin{thebibliography}{99}
\bibitem{Sinova} J. Sinova, S. O. Valenzuela, J. Wunderlich, C. H. Back, and
T. Jungwirth, Rev. Mod. Phys. \textbf{87}, 1213 (2015); M. Konig,
Science \textbf{318}, 766 (2007); Y. K. Kato, R. C. Myers, A. C. Gossard,
and D. D. Awschalom, \emph{ibid}. \textbf{306}, 1910 (2004).
\bibitem{Qi} X. L. Qi and S. C. Zhang, Physics Today \textbf{63}, 33 (2010);
 M. Z. Hasan and C. L. Kane, Rev. Mod. Phys. \textbf{82}, 3045 (2010)
\bibitem{Sau} J. D. Sau, R. M. Lutchyn, S. Tewari, and S. Das Sarma, Phys. Rev. Lett. \textbf{104}, 040502 (2010);
S.-L. Zhu, L.-B. Shao, Z. D. Wang, and L.-M. Duan, Phys. Rev. Lett.\textbf{ 106}, 100404 (2011).
\bibitem{Koralek} I. \u{z}uti\'{c},  J. Fabian, and  S. Das Sarma, Rev. Mod. Phys. \textbf{76}, 323 (2004); J. D. Koralek, C. P. Weber, J. Orenstein, B. A. Bernevig, S. C. Zhang, S. Mack, and D. D. Awschalom, Nature (London) \textbf{458}.
610 (2009);
\bibitem{Lin} Y.-J. Lin, K. Jim\'{e}nez-Garc\'{\i}a, and I. B. Spielman, Nature
(London) \textbf{471}, 83 (2011).
\bibitem{Pan1} J.-Y. Zhang, S.-C. Ji, Z. Chen, L. Zhang, Z.-D. Du, B. Yan, G.-S. Pan, B. Zhao, Y.-J. Deng, H. Zhai, S. Chen, and J.-W. Pan, Phys.
Rev. Lett. \textbf{109}, 115301 (2012).
\bibitem{Wang}  P. J. Wang, Z.-Q. Yu, Z. K. Fu, J. Miao, L. H. Huang, S. J. Chai,
H. Zhai, and J. Zhang, Phys. Rev. Lett. \textbf{109}, 095301 (2012).
\bibitem{Cheuk} L. W. Cheuk, A. T. Sommer, Z. Hadzibabic, T. Yefsah, W. S.
Bakr, and M. W. Zwierlein, Phys. Rev. Lett. \textbf{109}, 095302 (2012).
\bibitem{Huang} L. Huang, Z. Meng, P. Wang, P. Peng, S. L. Zhang, L.
Chen, D. Li, Q. Zhou, and J. Zhang, Nat. Phys. \textbf{12}, 540 (2016).
\bibitem{Pan}Z. Wu, L. Zhang, W. Sun, X.-T. Xu, B.-Z. Wang,
S.-C. Ji, Y. Deng, S. Chen, X.-J. Liu, and J.-W. Pan, Science \textbf{ 354}, 83 (2016).
\bibitem{Zhai} H. Zhai, Int. J. Mod. Phys. B \textbf{26}, 1230001 (2012); H. Zhai, Rep. Prog. Phys.\textbf{ 78}, 026001 (2015).
\bibitem{Wu} C. J. Wu, I. Mondragon-Shem, and X.-F. Zhou, Chin.
Phys. Lett. \textbf{28}, 097102 (2011);  X. F. Zhou, Y. Li, Z. Cai, and C. J. Wu, J. Phys. B: At. Mol. Opt.
Phys. \textbf{46}, 134001 (2013).
\bibitem{Stanescu} T. D. Stanescu, B. Anderson, and V. Galitski, Phys. Rev. A
\textbf{78}, 023616 (2008).
\bibitem{Sinha} S. Sinha, R. Nath, and L. Santos, Phys. Rev. Lett. \textbf{107},
270401 (2011).
\bibitem{Zhai2} C. J. Wang, C. Gao, C.-M. Jian, and H. Zhai, Phys. Rev. Lett.
\textbf{105}, 160403 (2010).
\bibitem{Ho} T.-L. Ho and S. Z. Zhang, Phys. Rev. Lett. \textbf{107}, 150403 (2011).
\bibitem{Li} Y. Li, L. P. Pitaevskii, and S. Stringari, Phys. Rev. Lett. \textbf{108},
225301 (2012).
\bibitem{Hu} H. Hu, B. Ramachandhran, H. Pu, and X.-J. Liu, Phys.
Rev. Lett. \textbf{108}, 010402 (2012).
\bibitem{Wen} L. Wen,  Q. Sun,  H. Q. Wang,  A. C. Ji, and W. M. Liu,  Phys. Rev. A \textbf{86}, 043602 (2012).
\bibitem{Zhang} Y. P. Zhang, L. Mao, and C. W. Zhang, Phys. Rev. Lett. \textbf{108},
035302 (2012).
\bibitem{Cole} W. S. Cole, S. Zhang, A. Paramekanti, and N. Trivedi,
Phys. Rev. Lett. \textbf{109}, 085302 (2012).
\bibitem{Radic} J. Radi\'{c}, A. Di Ciolo, K. Sun, and V. Galitski, Phys. Rev. Lett.
\textbf{109}, 085303 (2012).
\bibitem{Gong} M. Gong, Y. Qian, V. W. Scarola, and C. Zhang, Sci. Rep. \textbf{5}, 10050 (2015).
\bibitem{Cai} Z. Cai, X. Zhou, and C. Wu, Phys. Rev. A \textbf{85}, 061605(R) (2012).
\bibitem{Zhu1}D.-W. Zhang, J.-P. Chen, C.-J. Shan, Z. D. Wang, and S.-L. Zhu, Phys. Rev. A \textbf{ 88}, 013612 (2013).
\bibitem{Xu} Z. Xu, W. S. Cole, and S. Zhang, Phys. Rev. A \textbf{89}, 051604(R) (2014).
\bibitem{Pan3} J. Y. Zhang, S. C. Ji, Z. Chen, L. Zhang, Z. D. Du, B. Yan,
G. S. Pan, B. Zhao, Y. J. Deng, H. Zhai, S. Chen, and J. W. Pan,
Phys. Rev. Lett. \textbf{109}, 115301 (2012).
\bibitem{Zhai3}Z. Chen and H. Zhai Phys. Rev. A 86, 041604(R) (2012)
\bibitem{Qu} C. Qu, C. Hamner, M. Gong, C. W. Zhang, and P. Engels, Phys.
Rev. A \textbf{88}, 021604(R) (2013).
\bibitem{zhu2} D. W. Zhang, Z. Y. Xue, H. Yan, Z. D. Wang, and S. L. Zhu, Phys. Rev. A \textbf{85}, 013628 (2012).
\bibitem{zhu3} D. W. Zhang, L. B. Fu, Z. D. Wang, and S. L. Zhu, Phys. Rev. A
\textbf{85}, 043609 (2012).
\bibitem{Garcia-March} M. A. Garcia-March, G. Mazzarella, L. Dell'Anna, B. Juli\'{a}-D\'{\i}az, L. Salasnich, and A. Polls, Phys. Rev. A 89, 063607
(2014)
\bibitem{Citro} R. Citro and A. Naddeo, Eur. Phys. J. Spec. Top. 224, 503
(2015)
\bibitem{Xue} Z. F. Yu and J. K. Xue, Phys. Rev. A \textbf{90}, 033618 (2014).
\bibitem{Hai1}Y. Luo, G. Lu, C. Kong, and W. Hai, Phys. Rev. A\textbf{ 93}, 043409 (2016)
\bibitem{Ng} H. T. Ng, Phys. Rev. A \textbf{92}, 043634 (2015).
\bibitem{Olson} A. J. Olson, S.-J. Wang, R. J. Niffenegger, C.-H. Li, C. H. Greene, and Y. P. Chen, Phys. Rev. A \textbf{90}, 013616 (2014).
\bibitem{Larson}J. Larson, B. M. Anderson, and A. Altland, Phys. Rev. A \textbf{87}, 013624 (2013).
\bibitem{LZhou} L. Zhou, H. Pu, and W. Zhang, Phys. Rev. A \textbf{87}, 023625 (2013).
\bibitem{Adhikari}Y. Cheng, G. Tang, and S. K. Adhikari, Phys. Rev. A  \textbf{89}, 063602 (2014)
\bibitem{Edmonds} M. J. Edmonds, J. Otterbach, R. G. Unanyan, M. Fleischhauer,
M. Titov, and P.\"{o}hberg, New J. Phys. \textbf{14}, 073056 (2012).
\bibitem{Mardonov} Sh. Mardonov, M. Modugno, and E. Ya. Sherman, Phys. Rev.
Lett. \textbf{115,} 180402 (2015).
\bibitem{Hai2} C. Kong, H. Chen, C. Li, and W. Hai, CHAOS \textbf{28}, 023115 (2018).
\bibitem{Sun} F. X. Sun, W. Zhang, Q. Y. He, and Q. H. Gong, Phys. Rev. A  \textbf{97}, 012307 (2018).
\bibitem{Hamner}C. Hamner, Y. Zhang, M. A. Khamehchi, M. J. Davis, and P. Engels, Phys. Rev. Lett. \textbf{114}, 070401 (2015).
\bibitem{JMZhang}J. M. Zhang, D.Braak, and M. Kollar, Phys. Rev. Lett. \textbf{109}, 116405 (2012).
\bibitem{Zhong} H. Zhong, Z. Zhou, B. Zhu, Y.-G. Ke, and C. Lee, Chin. Phys. Lett. \textbf{34}, 070304 (2017).
\bibitem{Liang}A.-Z. Zhang, P. Zhang, D. Suqing, X.-G. Zhao, and J.-Q. Liang, Phys. Rev. B \textbf{63}, 045319 (2001).
\bibitem{Longhi} S. Longhi and G. Della Valle, Phys. Rev. A \textbf{86}, 042104 (2012).
\bibitem{Zhou1}Z. Zhou, S. Tang, H. Zhong, B. Zhu, Z. He, and J. Tan, J. Phys. B: At. Mol. Opt. Phys. \textbf{50}, 225002 (2017).
\bibitem{Bloch}Y. A. Chen, S. Nascimb\`{e}ne, M. Aidelsburger, M. Atala, S.
Trotzky, and I. Bloch, Phys. Rev. Lett. \textbf{107}, 210405 (2011).
\bibitem{Dunlap}D. H. Dunlap and V. M. Kenkre, Phys. Rev. B \textbf{34}, 3625 (1986).
\bibitem{Grossmann} F. Grossmann, T. Dittrich, P. Jung, and P. Hanggi, Phys.
Rev. Lett. \textbf{67}, 516 (1991); Z. Phys. B \textbf{84}, 315 (1991).
\bibitem{Longhi2} S. Longhi, Phys. Rev. B \textbf{77}, 195326 (2008).
\bibitem{Zhou2} Z. Zhou, W. Hai, Q. Xie, and J. Tan, New J. Phys. \textbf{15}, 123020 (2013).
\bibitem{Luo} X. B. Luo, D. Wu, S. Luo, Y. Guo,
X. Yu, and Q. Hu, J. Phys. A: Math. Theor \textbf{ 47}, 34530  (2014).
\bibitem{Spielman} K. Jim\'{e}nez-Garc\'{\i}a, L. J. LeBlanc, R. A. Williams, M. C. Beeler,
C. Qu, M. Gong, C. Zhang, and I. B. Spielman, Phys. Rev. Lett.
\textbf{114}, 125301 (2015).
\bibitem{You} X. Luo, L.Wu, J. Chen, Q. Guan, K. Gao, Z.-F. Xu, L. You, and
R. Wang, Sci. Rep. \textbf{6}, 18983 (2016).
\bibitem{YZhang}Y. Zhang, G. Chen, and C. Zhang, Sci. Rep. \textbf{ 3}, 1937 (2013).
\end{thebibliography}


\end{document}